%% file: main.tex
\documentclass[%
reprint,
 amsmath,amssymb,
 aps,
]{revtex4-2}

\usepackage{amsmath,amssymb, bm}
\usepackage{graphicx}
\usepackage{dcolumn}
\usepackage{bm}
\usepackage{physics}
\usepackage{comment}
\usepackage{ulem}
\usepackage{xcolor}
\usepackage{hyperref}

\usepackage{xifthen}
\newcommand{\vq}[2][]{                
  \ifthenelse{\isempty{#1}}           %
    { \hat{\pmb{#2}} }                
    { \hat{\pmb{#2}}_\mathrm{#1} }    
}
\newcommand{\vqtil}[2][]{             
  \ifthenelse{\isempty{#1}}           %
    { \hat{\til{\pmb{#2}}} }                
    { \hat{\til{\pmb{#2}}}_\mathrm{#1} }    
}
\newcommand{\tq}[2][]{                
  \ifthenelse{\isempty{#1}}           %
    { \mathbb{#2} }                   
    { \mathbb{#2}_\mathrm{#1} }       
}
\newcommand{\vs}[2][]{                
  \ifthenelse{\isempty{#1}}           %
    { \mathbf{#2} }                   
    { \mathbf{#2}_\mathrm{#1} }       
}
\newcommand{\matr}[4]{
  \begin{bmatrix}
    #1 & #2 \\
    #3 & #4
  \end{bmatrix}
}

\newcommand{\vect}[2]{
  \begin{bmatrix}
    #1  \\
    #2 
  \end{bmatrix}
}

\newcommand{\Tq}[3][]{                
  \ifthenelse{\isempty{#1}}           %
    { \mathbb{#3} }                   
    { \mathbb{#3}^{\rm #1}_{\rm #2} }       
}

\newcommand{\mr}[1]{\mathrm{#1}}

\date{\today}

\begin{document}
\preprint{APS/123-QED}
\title{Teleportation-based filtering for gravitational-wave detectors}
\author{Yohei Nishino$^{1,2}$}
 \email{yohei.nishino@grad.nao.ac.jp}

\affiliation{$^1$Department of Astronomy, University of Tokyo, Bunkyo, Tokyo 113-0033, Japan,}
\affiliation{$^2$Gravitational Wave Science Project, National Astronomical Observatory of Japan (NAOJ), Mitaka City, Tokyo
181-8588, Japan,}

\date{\today}

\begin{abstract}
Employing the principle of quantum teleportation in gravitational-wave detectors provides a method to address technical challenges associated with implementing quantum filters for squeezed-light injection. A recent proposal [Phys.\ Rev.\ A \textbf{110}, 022601] has shown that an arbitrary number of effective squeezing-angle rotations can be achieved by cascading quantum teleportation, yielding sensitivity equivalent to frequency-dependent pre-filtering. A natural question is whether the teleportation-based protocol can also realize frequency-dependent post-filtering. In this work, we examine this possibility and show that, with perfect teleportation fidelity, post-filtering can be fully implemented. In practice, however, various imperfections make the sensitivity nearly equivalent to that of teleportation-based pre-filtering. We apply this scheme to the low-frequency detector in the Einstein Telescope xylophone configuration and find that teleportation-based post-filtering could also be a candidate for the detector’s design.
\end{abstract}

\maketitle
\input{chap_main}

\bibliography{ref} %

\end{document}

%% file: chap_main.tex
\section{Introduction}
The first direct detection of gravitational waves (GW) from a binary black hole merger in 2015 by the two LIGO detectors marked the beginning of gravitational-wave astronomy~\cite{PhysRevLett.116.061102}. Since then, numerous merger events have been observed, providing insights into astrophysical phenomena~\cite{PhysRevX.9.031040,PhysRevX.11.021053,PhysRevD.109.022001,PhysRevX.13.041039}. The ongoing development of current-generation detectors such as LIGO~\cite{Aasi_2015}, Virgo~\cite{Acernese_2015}, and KAGRA~\cite{Somiya_2012}, as well as the planned next-generation Einstein Telescope~\cite{Hild_2011} and Cosmic Explorer~\cite{Abbott_2017}, is expected to extend the reach of gravitational-wave observations. These instruments will allow more precise tests of general relativity in strong-field regimes~\cite{Will_2014, Berti_2015}, the determination of neutron star equations of state~\cite{LATTIMER_2007}, and studies of the early universe~\cite{Sathyaprakash_2009,Caprini_2018}.

Gravitational-wave detectors are highly-sensitive weak-force sensors, measuring the tidal forces exerted by GWs on suspended test masses (mirrors) using light as the readout medium~\cite{1995qume.book.....B}. Quantum noise in the light field---shot noise and radiation-pressure noise---imposes fundamental limits on sensitivity. Due to the Heisenberg uncertainty principle, these two noise components are coupled by a trade-off known as the standard quantum limit (SQL)~\cite{1968JETP.26.831,PhysRevLett.45.75}. A common method to surpass the SQL in current detectors involves frequency-dependent rotation of the squeezing angle using filter cavities~\cite{PhysRevD.65.022002}. In this approach, either the input or the output optical field passes through a detuned Fabry–P\'erot cavity to compensate the frequency-dependent rotation of the optomechanical squeezing angle in the interferometer.

In this paper, we refer to filtering of the interferometer's input field as \textit{pre-filtering} and filtering of the output field as \textit{post-filtering}. The concept of frequency-dependent filtering has developed through several stages. Early proposals considered the injection of squeezed vacuum states generated via parametric down-conversion~\cite{Meystre1983QuantumOE,PhysRevD.23.1693} and the use of variational readout techniques~\cite{VYATCHANIN1993772218,VYATCHANIN1994492,VYATCHANIN1995269,VYATCHANIN19968261007,VYATCHANIN1996834690}. Kimble et al.\ (2001) introduced the use of Fabry–P\'erot cavities to filter squeezed vacuum before injection into a tuned interferometer, and also to filter ponderomotively squeezed vacuum at the output~\cite{PhysRevD.65.022002}. Their scheme initially required two filter cavities. Purdue and Chen developed a more general framework based on interferometer response functions~\cite{PhysRevD.66.122004} and it is shown later that a single filter cavity is sufficient for broadband tuned interferometers~\cite{Danilishin_2019}. Buonanno and Chen subsequently demonstrated that detuned interferometers could surpass the SQL~\cite{PhysRevD.64.042006, PhysRevD.65.042001,PhysRevD.67.062002}, and Harms et al.\ calculated the optimal performance of frequency-dependent pre-filtering for such configurations~\cite{PhysRevD.68.042001}. Buonanno and Chen also optimized post-filtering for detuned interferometers~\cite{PhysRevD.69.102004}. Khalili (2010) included optical losses in the analysis for tuned interferometers, concluding that Kimble’s pre-filtering remained optimal because post-filtering modifies the detection angle and is more susceptible to the interferometer loss~\cite{PhysRevD.81.122002}.

Due to these considerations, pre-filtering has become the dominant approach in both current and future gravitational-wave detectors, regardless of whether the interferometer is tuned or detuned. To enhance the performance of filter cavities, increasing their length is advantageous~\cite{PhysRevD.81.122002}. Current-generation detectors employ filter cavities on the order of hundreds of meters~\cite{PhysRevX.13.041021,PhysRevLett.131.041403}, while next-generation facilities plan to extend these cavities to several kilometers~\cite{galaxies10040090,galaxies13010011}. However, constructing kilometer-scale filter cavities presents financial and infrastructural challenges.

An alternative approach to mitigating this issue involves leveraging quantum entanglement for filtering. Ma et al.\ (2017) proposed conditional filtering for tuned interferometers~\cite{Ma_2017}, a scheme that can be interpreted as either pre-filtering or post-filtering. More recently, Nishino et al.\ (2024) introduced teleportation-based pre-filtering for detuned interferometers~\cite{PhysRevA.110.022601}. These methods repurpose the interferometer itself as a filter cavity by utilizing its long baseline, low loss, and stable optical interferometer, thereby avoiding the construction of additional filter cavities and reducing costs associated with infrastructure.

In this paper, we revisit the protocol and investigate the feasibility of teleportation-based post-filtering. The main claim is that teleportation-based post-filtering yields better sensitivity for detuned interferometers than teleportation-based pre-filtering in the limit of perfect teleportation fidelity. Under realistic parameter regimes, however, we find that imperfections in the teleportation process and optical losses diminish the advantages of post-filtering. Using the parameters of the Einstein Telescope, we find that post-filtering achieves nearly the same sensitivity as pre-filtering.

The rest of the paper is structured as follows: In Sec.~\ref{sec.2}, we briefly review teleportation-based filtering schemes. In Sec.~\ref{sec.3}, we present a semi-analytical formalism for teleportation-based filtering. In Sec.~\ref{sec.4}, we apply the teleportation protocol to a detuned interferometer, and in Sec.~\ref{sec.5}, we evaluate the resulting sensitivity in the Einstein Telescope. In Sec.~\ref{sec.6}, we compare pre-filtering and post-filtering quantitatively. Finally, we conclude in Sec.~\ref{sec.7}.

\begin{figure*}[ht]
    \centering
    \includegraphics[width=0.95\linewidth]{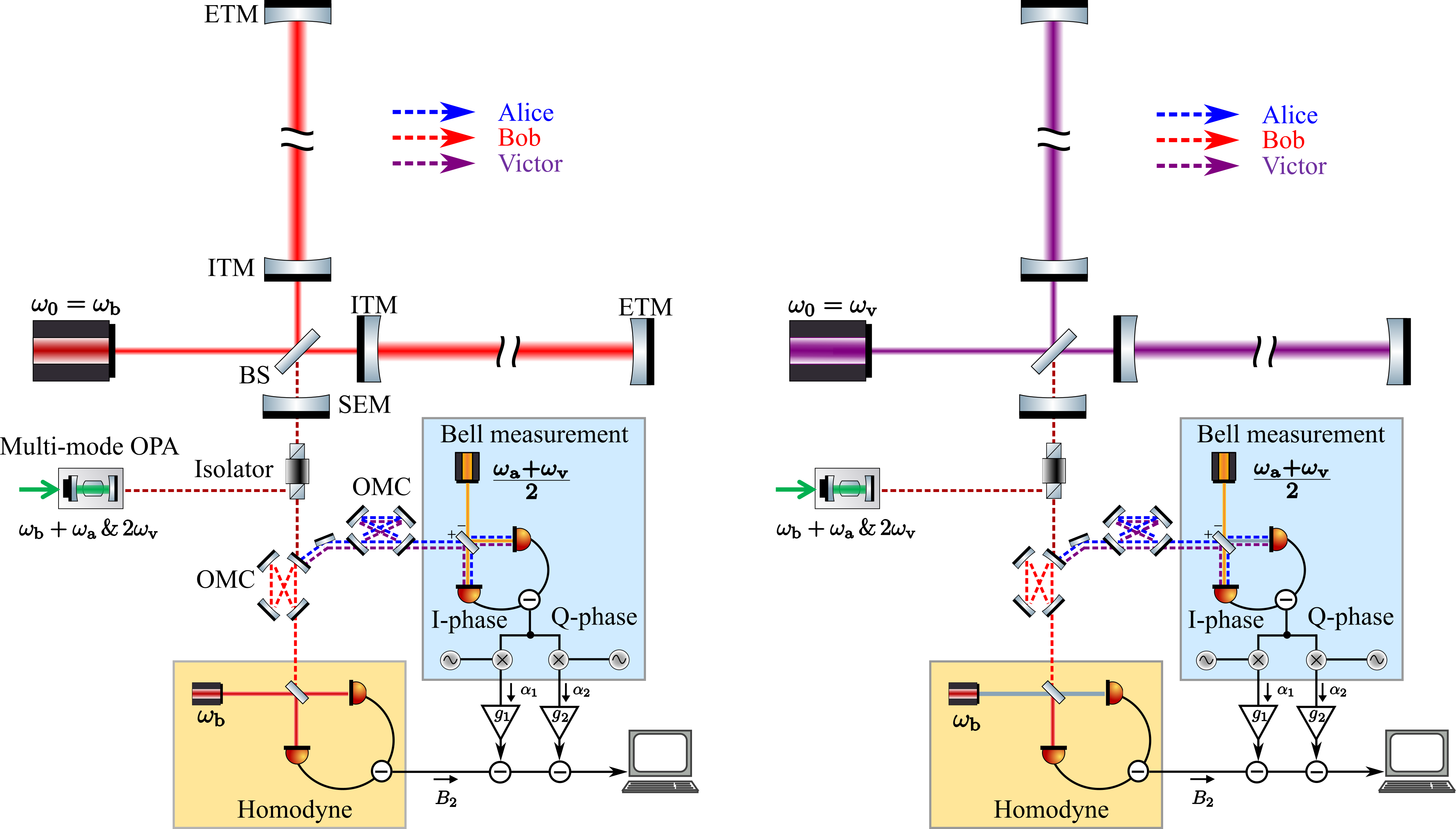}
    \caption{Detector configuration for teleportation-based pre-filtering (left) and post-filtering (right). The OPA at the dark port is pumped at two frequencies, $\omega_\mr{b}+\omega_\mr{a}$ and $\omega_\mr{v}$. Three fields at different frequencies are injected into the detector through the isolator. The outgoing fields are separated by the output mode cleaner. Bob's field is sent to the homodyne detector, while Victor's and Alice's fields are sent to the Bell measurement. The measurement outputs are combined with optimal filters $g_1$ and $g_2$, and Victor's state is teleported to Bob. Abbreviations: ITM, input test mass; ETM, end test mass; BS, beam splitter; NOPA, non-degenerate optical parametric amplifier; SEM, signal extraction mirror; OMC, output mode cleaner.}
    \label{fig:Fig_two_configuration}
\end{figure*}

\section{Review of Teleportation-Based Filtering}\label{sec.2}
A central motivation for teleportation-based filtering is to reduce the infrastructure-induced costs of long-baseline detectors by re-purposing the detector itself as a filter cavity. A Fabry–P\'erot Michelson interferometer, the standard configuration for gravitational-wave detectors, can be decomposed into common and differential modes. From the perspective of the anti-symmetric port, where gravitational-wave-induced sidebands exit, the two cavities and the Michelson interferometer can effectively be regarded as a single cavity.

Since a single cavity supports multiple frequency modes, each separated by the free spectral range, it is possible to engineer distinct responses for different frequencies within the single-mode approximation. Furthermore, a three-mirror cavity composed of the signal extraction mirror, input test mass, and end test mass (see Fig.~\ref{fig:Fig_two_configuration}) can exhibit different bandwidths for different frequency modes. Conditional squeezing~\cite{Ma_2017} and teleportation-based squeezing~\cite{PhysRevA.110.022601} exploit this property to achieve frequency-dependent phase rotation. In other words, the signal extraction mirror is inherently
necessary for these schemes.

\begin{figure}[h]
    \centering
    \includegraphics[width=1\linewidth]{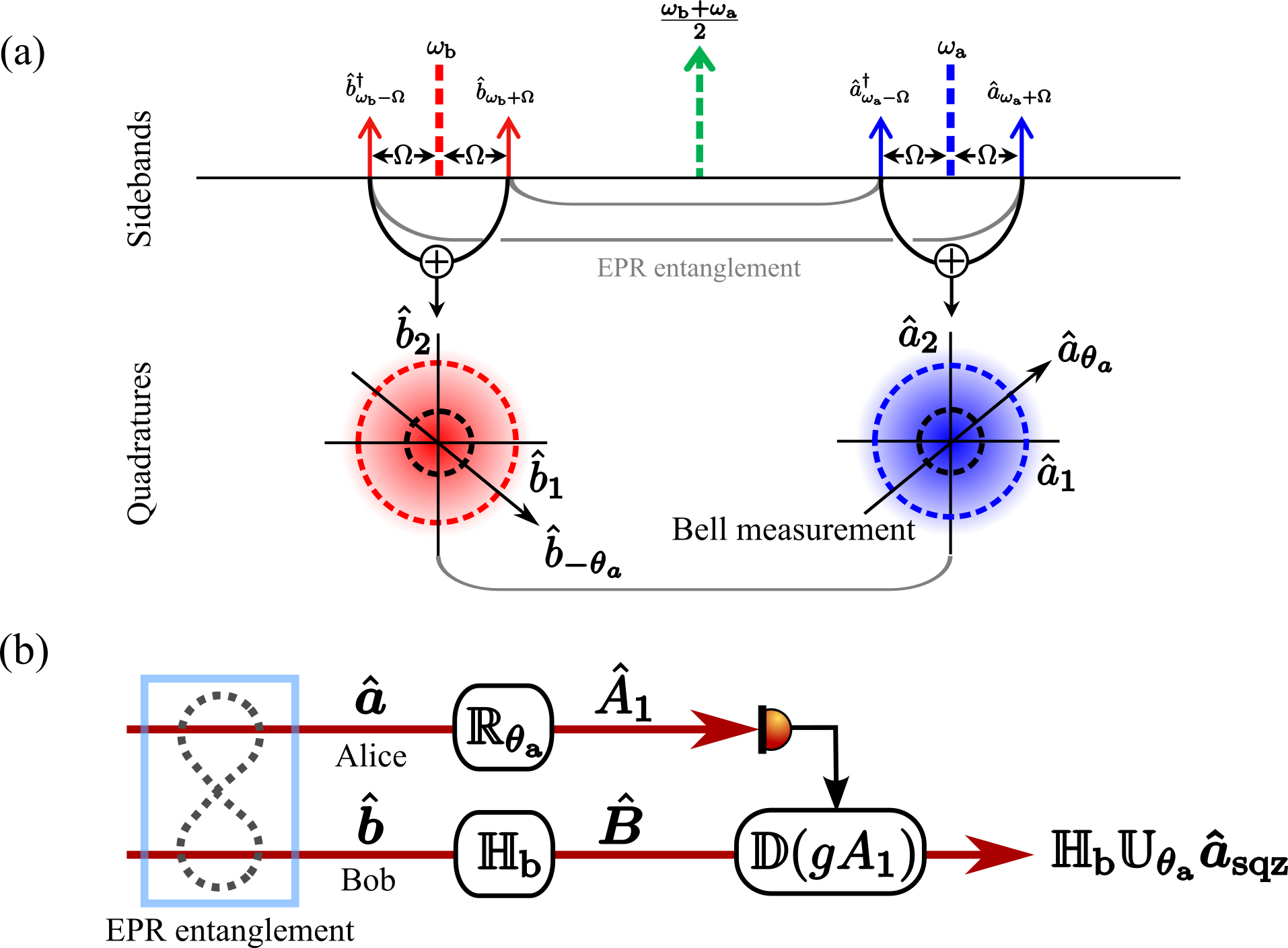}
    \caption{(a) Fields of Alice and Bob in the sideband and quadrature pictures. (b) Schematic of the conditional squeezing. Alice and Bob are entangled; Alice undergoes a phase rotation ($\tq[\theta_\mr{a}]{R}$), and Bob undergoes ponderomotive squeezing ($\tq[\mr{b}]{H}$). Upon detection of Alice, Bob is conditionally squeezed along a specific quadrature. Displacement (or post-processing) according to Alice's measurement outcome conditions Bob's state, and the final output becomes equivalent to conventional pre-filtering with a single filter cavity.}
    \label{fig:Fig_EPR_squeezing}
\end{figure}

It is important to note that conditional squeezing can be interpreted as sideband teleportation, as presented in Ref.~\cite{Chen2016} (see Fig.~\ref{fig:Fig_EPR_squeezing}a). Entanglement is generated in the frequency domain using a non-degenerate optical parametric amplifier (NDOPA). The fields forming the entangled pair are called Alice and Bob, and the sideband fields are distributed around their central frequencies as \( \omega_{\mr{a,b}} \pm \Omega \). The upper sideband of Alice at \( \omega_\mr{a}+\Omega \) and the lower sideband of Bob at \( \omega_\mr{b}-\Omega \) are entangled because the pump frequency \( \omega_\mr{b}+\omega_\mr{a} \) equals their sum. Similarly, the lower sideband of Alice at \( \omega_\mr{a}-\Omega \) and the upper sideband of Bob at \( \omega_\mr{b}+\Omega \) are also entangled. Homodyne detection of Alice, which simultaneously measures the upper and lower sidebands, can be interpreted as a ``Bell" measurement of the lower sideband at \( \omega_\mr{a}-\Omega \), with the upper sideband at \( \omega_\mr{a}+\Omega \) serving as EPR noise. Since the upper sideband at \( \omega_\mr{a}+\Omega \) is entangled with the lower sideband at \( \omega_\mr{b}-\Omega \), the upper sideband of Alice is teleported to Bob’s lower sideband after displacement. As a result, Bob's field gets squeezed. Thus, we categorize conditional squeezing as a class of ``teleportation-based filtering".

Conditional squeezing can also be viewed as an imperfect post-filtering. Depending on Alice’s phase rotation, the squeezed quadrature of Bob before ponderomotive squeezing (\( \tq[b]{H} \) in Fig.~\ref{fig:Fig_EPR_squeezing}b) undergoes conditional phase rotation, making the output equivalent to pre-filtering. If the displacement operation is replaced with post-processing and the system is analyzed from Alice’s perspective, its state before the phase rotation is conditionally and ponderomotively squeezed upon Bob’s detection. After post-processing, the output becomes equivalent to frequency-dependent post-filtering. However, due to imperfections in the EPR entanglement, the low-frequency sensitivity remains limited by shot noise.

In the rest of the section, we outline several advantages and drawbacks of teleportation-based filtering.

\subsection*{Advantages}
We outline three advantages. One significant advantage is that physical filter cavities are not required. For underground detectors such as the Einstein Telescope (ET), constructing long-baseline filter cavities requires costs due to the need for vacuum chambers, caverns, control systems, and related infrastructure. In contrast, modifications required for the teleportation-based scheme are limited to the detection port, where the entangled state is generated and additional detection ports are implemented.

Using the detector itself as a filter cavity also reduces the impact of filter imperfections. In the case of the ET, the arm cavities are designed to be 10\,km long, and the round-trip loss in these cavities will be reduced to approximately 45\,ppm. Consequently, when filter performance is evaluated in terms of loss per unit length~\cite{PhysRevD.81.122002}, the effect of losses is less pronounced than in shorter filter cavities. Furthermore, the arm cavity length is stabilized by common-mode control and multi-stage suspensions, which suppresses phase noise caused by filter-cavity length fluctuations. This is the reason why the teleportation-based squeezing in Ref.~\cite{PhysRevA.110.022601} outperforms the conventional scheme at the optical-spring frequency.

In addition, the teleportation-based scheme is flexible with respect to changes in detector parameters. Since the filter bandwidth is determined by
\begin{align}
    \gamma = \frac{c\,T}{4\,L}, \label{eq:bandwidth}
\end{align}
where \(T\) is the power transmissivity of the cavity input mirror and \(L\) is the cavity length, the bandwidth is fixed once the input mirror is manufactured (assuming \(T\) remains constant). In the teleportation-based scheme, one only needs to adjust the macroscopic but short lengths of the arm and the signal extraction cavity (SEC) and to change the frequencies of the idler beams.

\subsection*{Drawbacks}
We outline three drawbacks. A main drawback is that injection and readout losses are amplified by a factor of three compared to conventional detection, as three beams are involved. As shown in later sections, the dominant noise contributions in the quantum-noise-limited sensitivity are the vacuum fluctuations induced by these losses. The loss budget considered in the analysis includes contributions from the OPA cavity, the Faraday isolator, and the photodetectors. 

Second, because three detection ports are used, a degradation of the squeezing level by a few decibels is unavoidable. This degradation can be compensated by increasing the squeezing generated by the OPA. For clarity, note that the squeezing level of 15\,dB refers to the \textit{generated} squeezing before any losses (including OPA losses), whereas the \textit{observed} squeezing level is approximately 5\,dB. This level of squeezing is achievable with current technology.

Finally, the need for macroscopic tuning of the arm and SEC lengths and the precise selection of the idler frequencies adds technical complexity during interferometer commissioning. The challenge arises because the idler frequencies are separated from the carrier frequency by radio frequencies (MHz to GHz), while the effective bandwidths and detunings of the filter cavities are on the order of $\sim10$\,Hz. A robust acquisition scheme for such tuning must be developed.

\section{Teleportation-Based Filtering}\label{sec.3}

\begin{figure}[]
    \centering
    \includegraphics[scale=0.65]{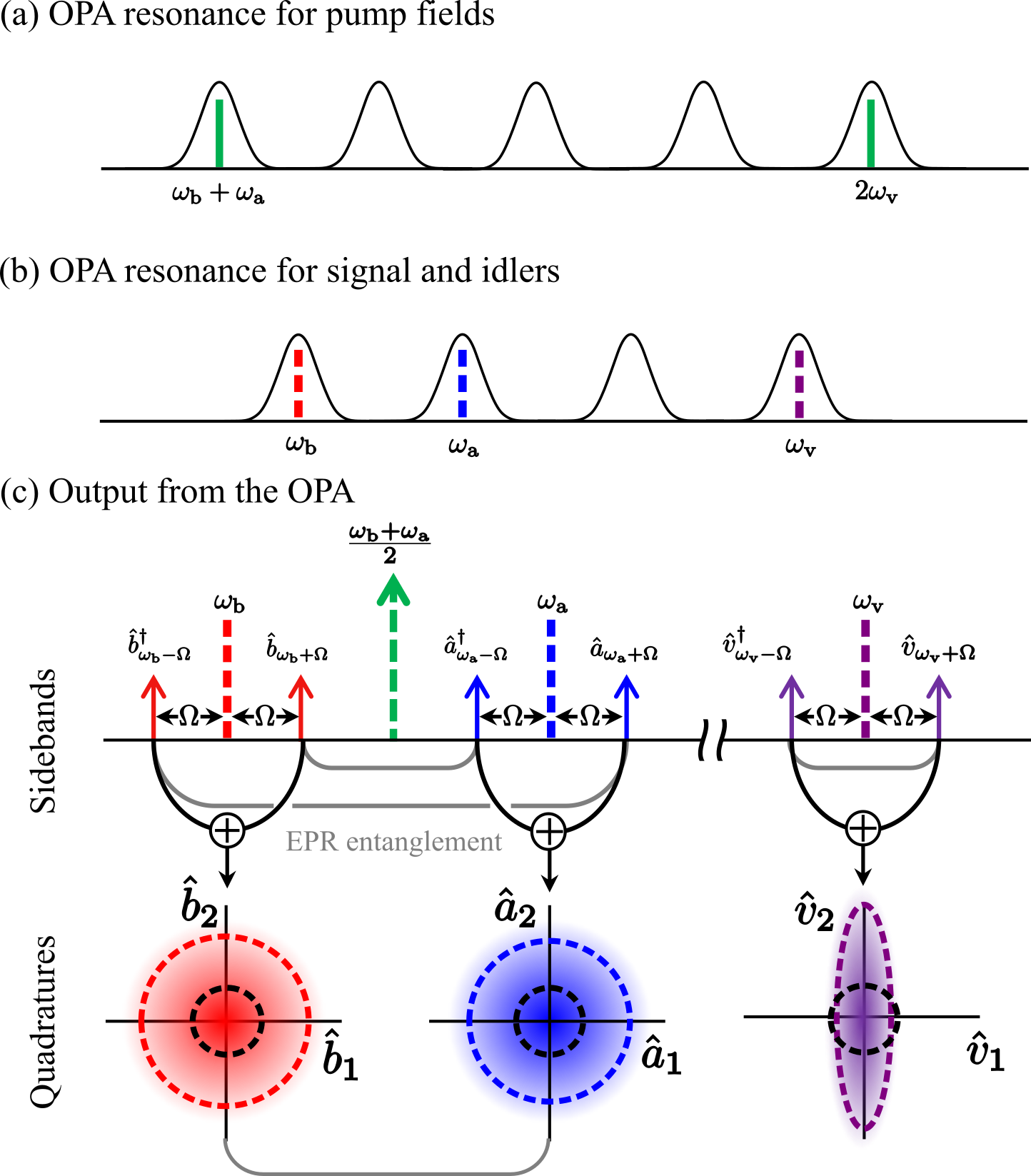}
    \caption{(a) OPA resonance for two pump fields, with frequency difference \(\omega_\mr{b}+\omega_\mr{a}-\omega_\mr{v}\). (b) Schematic of three beams. The top and bottom panels depict the fields in the sideband and quadrature representations, respectively. The OPA is pumped at two frequencies, \(\omega_\mr{b}+\omega_\mr{a}\) and \(2\omega_\mr{v}\), generating entanglement symmetrically at the sideband frequencies. The first pump produces EPR entanglement between Alice and Bob, while the second generates a squeezed state for Victor.}
    \label{fig:continous_fields}
\end{figure}

\begin{figure}[]
    \centering
    \includegraphics[width=0.95\linewidth]{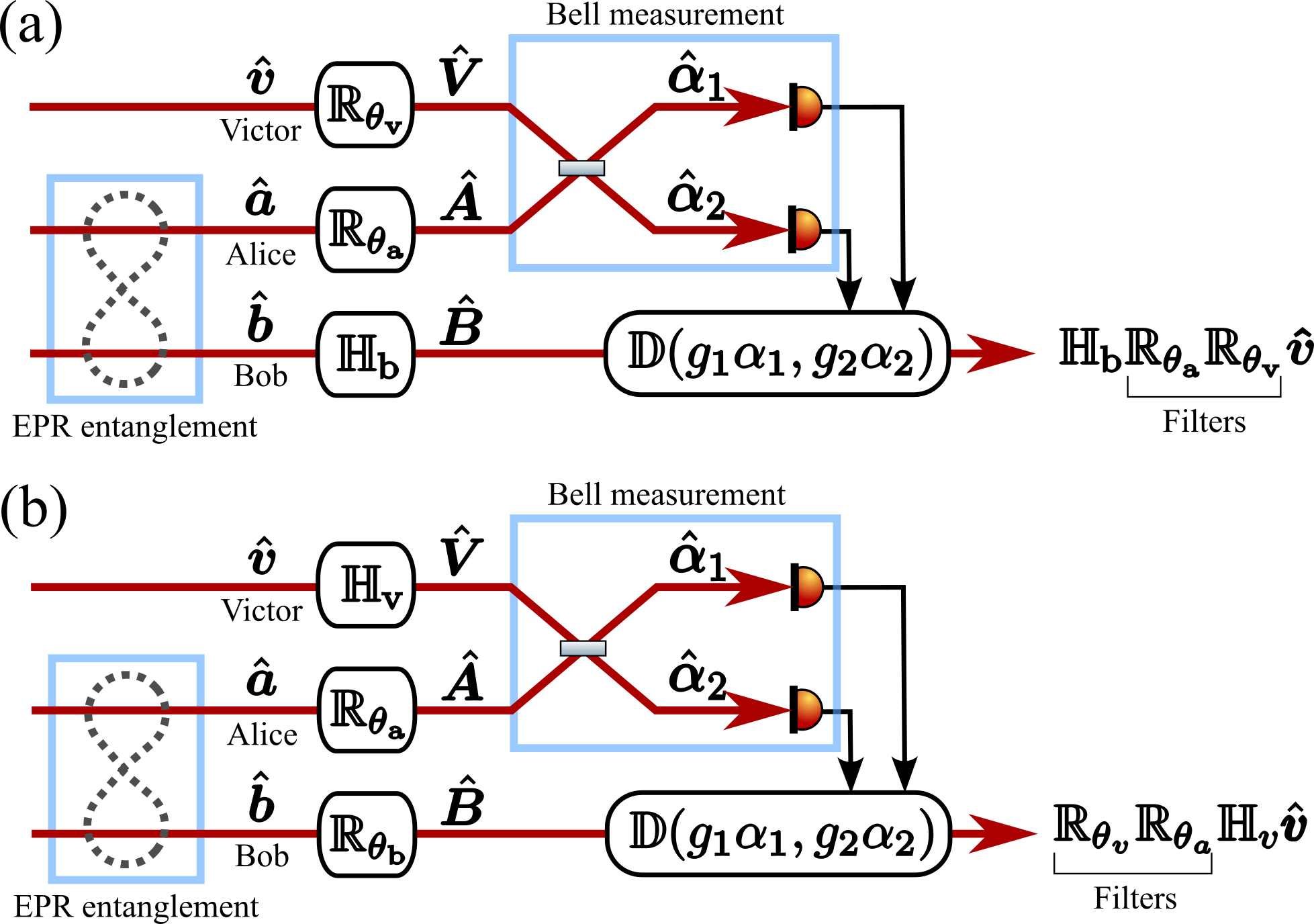}
    \caption{Schematic of teleportation-based (a) pre-filtering and (b) post-filtering. Alice and Bob are entangle in the EPR manner. By changing the order of phase rotations and ponderomotive squeezing, the operation mode can be switched from pre-filtering to post-filtering. Entanglement and Bell measurement are realized in the frequency domain, and the displacement operation is replaced with post-processing.}
    \label{fig:Fig_modes_teleportation}
\end{figure}

\subsection{Quantum State Preparation}\label{s.sec2.1}
The initial quantum states of the three beams—Alice, Bob, and Victor—are prepared using a multi-mode OPA. The entangled beams for Alice and Bob, centered at frequencies \(\omega_\mr{a}\) and \(\omega_\mr{b}\), are generated by driving the OPA with a pump beam at frequency \(\omega_\mr{b}+\omega_\mr{a}\). The squeezed beam for Victor, at frequency \(\omega_\mr{v}\), is produced by a second pump beam at frequency \(2\omega_\mr{v}\) (see Fig.~\ref{fig:continous_fields}a). The sidebands at $\pm\Omega$ around the three frequencies are assumed to lie within the OPA bandwidth, and \(\omega_\mr{v}-\omega_\mr{a}\) and \(\omega_\mr{a}-\omega_\mr{b}\) are integer multiples of the OPA cavity free spectral range (see Fig.~\ref{fig:continous_fields}b).

In the two-photon formalism~\cite{PhysRevA.31.3093,PhysRevA.31.3068}, the fluctuations of the light are described by a two-dimensional vector of amplitude and phase quadrature operators (see Fig.~\ref{fig:continous_fields}c). For Alice, Bob, and Victor, we define
\begin{align}
    \pmb{\hat{a}} &= \{\hat{a}_{1,\Omega},\,\hat{a}_{2,\Omega}\}^T,\quad
    \pmb{\hat{b}} = \{\hat{b}_{1,\Omega},\,\hat{b}_{2,\Omega}\}^T,\notag\\
    \pmb{\hat{v}} &= \{\hat{v}_{1,\Omega},\,\hat{v}_{2,\Omega}\}^T,
\end{align}
where \(\Omega\) denotes the sideband frequency and the superscript \(T\) indicates the transpose. For brevity, we omit the superscript \(\Omega\) in what follows.

Detection is described by projecting the output field onto the homodyne vector
\begin{align}
    \pmb{H}_{\zeta}^T = \{\cos\zeta,\,\sin\zeta\},
\end{align}
so that
\begin{align}
    \hat{a}_\zeta = \pmb{H}_{\zeta}^T\,\pmb{\hat{a}}_\mathrm{out},
\end{align}
with output field $\vq[out]{a}$. The quantum spectral density of the input fields, \(\tq[\mu]{S}^\mathrm{in}\), is defined by
\begin{align}
    2\pi\,\delta(\Omega-\Omega')\,\tq[\mathit{ij}]{S}^\mathrm{in}(\Omega)
    &\equiv \tfrac{1}{2}\langle \mathrm{in}\mid \hat{a}_{i,\mathrm{in}}(\Omega)\,\hat{a}_{j,\mathrm{in}}^\dagger(\Omega') \notag\\
    &\quad+ \hat{a}_{j,\mathrm{in}}^\dagger(\Omega')\,\hat{a}_{i,\mathrm{in}}(\Omega)\mid \mathrm{in}\rangle,
\end{align}
with \((i,j)=\{1,2\}\) (see Ref.~\cite{Danilishin_2012} for details). When the output field has a response matrix \(\tq{T}\) and a signal vector \(\vq{t}\), it is expressed in terms of the input $\vq[in]{a}$ as
\begin{align}
    \pmb{\hat{a}}_\mathrm{out} = \tq{T}\,\pmb{\hat{a}}_\mr{in} + \pmb{t}_h\,\frac{h}{h_\mathrm{SQL}}.
\end{align}
Accordingly, the force spectral density is
\begin{align}
    S_h(\Omega) = h_\mathrm{SQL}^2\,\frac{\sum_{\mu} \pmb{H}_{\zeta}^T\,\tq[]{T}\,\tq[]{S}^\mathrm{in}\,\tq[]{T}^\dagger\,\pmb{H}_{\zeta}}{\bigl|\pmb{H}_{\zeta}^T\,\pmb{t}_h\bigr|^2}. \label{eq:SF}
\end{align}

The strength of the EPR entanglement between Alice's and Bob's states can be characterized by the spectral densities of the four EPR operators~\cite{2000_Duan} \((\hat{a}_{i}\pm\hat{b}_{i})/\sqrt{2}\), such that
\begin{align}
    S_{(\hat{a}_{1}\pm\hat{b}_{1})/\sqrt{2}} &= e^{\pm 2r}, \quad
    S_{(\hat{a}_{2}\pm\hat{b}_{2})/\sqrt{2}} = e^{\mp 2r}, \label{eq:spectrum}
\end{align}
where \(r\) denotes the squeezing factor. In the limit \(r\rightarrow\infty\), the noise spectra \(S_{(\hat{a}_{1}-\hat{b}_{1})/\sqrt{2}}\) and \(S_{(\hat{a}_{2}+\hat{b}_{2})/\sqrt{2}}\) vanish, corresponding to ideal EPR entanglement~\cite{PhysRev.47.777}. More generally, when measuring the quadrature \(\hat{a}_{-\theta}=\hat{a}_1\cos\theta-\hat{a}_2\sin\theta\), the quadrature \(\hat{b}_{\theta}=\hat{b}_1\cos\theta+\hat{b}_2\sin\theta\) becomes conditionally squeezed, and vice versa. The spectral density of the conditionally squeezed field is
\begin{equation}\label{eq:Stheta}
S_{\hat{b}_{\theta}\hat{b}_{\theta}}^{\hat{a}_{-\theta}}=\frac{1}{\cosh(2r)}, \quad
S_{\hat{b}_{\pi/2+\theta}\hat{b}_{\pi/2+\theta}}^{\hat{a}_{-\theta}}=\cosh(2r).
\end{equation}
The amplitude (phase) quadrature of Victor exhibits squeezing (anti-squeezing) as
\begin{align}
    S_{\hat{v}_1\hat{v}_1}=e^{-2r_v}, \quad S_{\hat{v}_2\hat{v}_2}=e^{2r_v}.
\end{align}

For later discussion, it is useful to introduce the Pauli matrix~\cite{nishino2025teleportationbasedspeedmeterprecision}:
\begin{align}
    \tq[]{M} = \matr{-1}{0}{0}{1},
\end{align}
with which the Bell observables can be expressed in vector form as
\begin{align}
    \pmb{\hat{\alpha}} = \frac{\vq[]{V} + \tq[]{M}\vq[]{A}}{\sqrt{2}}
    = \frac{1}{\sqrt{2}}
    \begin{pmatrix}
        \hat{V}_1-\hat{A}_1 \\
        \hat{V}_2+\hat{A}_2
    \end{pmatrix}.
\end{align}
Similarly, the noise-suppressed quadratures in Eq.~\eqref{eq:spectrum} can be written using \(\tq[]{M}\) as
\begin{align}
    \pmb{\hat{z}} = \frac{\pmb{\hat{b}}+\tq{M}\pmb{\hat{a}}}{\sqrt{2}}
    = \frac{1}{\sqrt{2}}
    \begin{pmatrix}
        \hat{b}_1-\hat{a}_1 \\
        \hat{b}_2+\hat{a}_2
    \end{pmatrix},
\end{align}
with covariance matrix
\begin{align}
    \tq{S}_{\hat{z}} = \tfrac{1}{2}
    \matr{e^{-2r}}{0}{0}{e^{-2r}}. \label{eq:S_z}
\end{align}
Equation~\eqref{eq:S_z} converges to a null matrix as \(r\rightarrow\infty\).

\subsection{Noise Filtering via Teleportation}
We next discuss the interferometer response and the detection strategy for teleportation. The teleportation procedure is described following Ref.~\cite{FURUSAWA200797}. A more detailed analysis is presented in a later section.

The input fields \(\vq[]{v}\), \(\vq[]{a}\), and \(\vq[]{b}\) pass through the interferometer, whose response is characterized by a symplectic transformation \(\tq[]{H}\) and displacement vector \(\vq[]{t}\):
\begin{align}
    \vq[]{V} &= \tq[v]{H} \vq[]{v} + \vq[v]{t}\,\frac{h}{h_\mathrm{SQL}}, \notag\\
    \vq[]{A} &= \tq[a]{H} \vq[]{a} + \vq[a]{t}\,\frac{h}{h_\mathrm{SQL}}, \notag\\
    \vq[]{B} &= \tq[b]{H} \vq[]{b} + \vq[b]{t}\,\frac{h}{h_\mathrm{SQL}}.
\end{align}
Bob's output before detection can be decomposed as
\begin{align}
    \vq[]{B} &= \underbrace{\tq[b]{H} \sqrt{2}\,\vq[]{z}}_{\text{EPR noise}}
    + \underbrace{\tq[b]{H} \tq[]{M} \tq[a]{H}^{-1} \sqrt{2}\,\vq[]{\alpha}}_{\text{Bell observable}} \notag\\
    &\quad + \tq[b]{H} \tq[]{M} \tq[a]{H}^{-1} \tq[]{M} \tq[v]{H}\vq[]{v} \notag\\
    &\quad + \frac{h}{h_\mathrm{SQL}} \Bigl[\vq[b]{t}
      + \tq[b]{H} \tq[]{M} \tq[a]{H}^{-1} \vq[a]{t}
      + \tq[b]{H} \tq[]{M} \tq[a]{H}^{-1} \tq[]{M} \tq[v]{H}\vq[v]{t}\Bigr], \label{eq:B}
\end{align}
where the following substitutions have been used:
\begin{align}
    \vq[]{b} &\rightarrow \sqrt{2}\,\vq[]{z} + \tq{M}\pmb{\hat{a}}, \\
    \vq[]{a} &\rightarrow \tq[a]{H}^{-1}\!\left[\vq[]{A}-\vq[a]{t}\,\frac{h}{h_\mathrm{SQL}}\right], \\
    \vq[]{A} &\rightarrow -\vq[]{\alpha} - \tq[]{M}\vq[]{V}.
\end{align}

The first term in Eq.~\eqref{eq:B} corresponds to the EPR noise, whose covariance is given by Eq.~\eqref{eq:S_z}; with ideal EPR entanglement its contribution vanishes. In the ideal case, the teleportation fidelity
\begin{align}
    F = \frac{1}{1+e^{-2r}}
\end{align}
approaches unity as \(r\rightarrow\infty\)~\cite{doi:10.1126/science.282.5389.706}. In practice, imperfections in the entanglement introduce additional shot noise.

Upon the Bell measurement, the observables \(\hat{\alpha}_1\) and \(\hat{\alpha}_2\) collapse to classical values, i.e., \(\vq[]{\alpha} \rightarrow \pmb{\alpha}\). Consequently, the second term in Eq.~\eqref{eq:B} can be eliminated through displacement (or post-processing). In the limit \(r\rightarrow\infty\), Bob's output reduces to
\begin{align}
    \vq[]{B} &\rightarrow \tq[b]{H} \tq[]{M} \tq[a]{H}^{-1} \tq[]{M} \tq[v]{H} \vq[]{v} \notag\\
    &\quad + \frac{h}{h_\mathrm{SQL}} \Bigl[\vq[b]{t}
      + \tq[b]{H} \tq[]{M} \tq[a]{H}^{-1} \vq[a]{t}
      + \tq[b]{H} \tq[]{M} \tq[a]{H}^{-1} \tq[]{M}\vq[v]{t}\Bigr],
\end{align}
which represents the general form of the teleportation output. In principle, arbitrary symplectic transformations may be implemented on the three fields (e.g., by pumping all three modes), with the signal-to-noise ratio optimized according to an appropriate figure of merit. In the following, we restrict the analysis to the case of a single pump and examine how the output can be exploited for filtering quantum noise.

\subsection{Pre-Filtering}
We show that the scheme reduces to the pre-filtering approach presented in Ref.~\cite{PhysRevA.110.022601} when only Bob's mode is pumped, i.e., when $\omega_\mr{b}=\omega_0$ and \(\vq[v,a]{t}=\{0,0\}^T\) (see Fig.~\ref{fig:Fig_modes_teleportation}a). In this case, the output simplifies to
\begin{align}
    \vq[]{B} \rightarrow \tq[b]{H} \tq[]{M} \tq[a]{H}^{-1} \tq[]{M} \tq[v]{H} \vq[]{v}
    + \frac{h}{h_\mathrm{SQL}} \vq[b]{t}.
\end{align}
Assuming that the operations for Victor and Alice are phase rotations,
\begin{align}
    \tq[v,a]{H} = \tq[\theta_{v,a}]{R}
    = \matr{\cos\theta_{v,a}}{-\sin\theta_{v,a}}
           {\sin\theta_{v,a}}{\cos\theta_{v,a}},
\end{align}
and that Bob's operation is a shear matrix corresponding to ponderomotive squeezing, the output becomes
\begin{align}
    \vq[]{B} \rightarrow \tq[b]{H}\,\tq[\theta_{a}]{R}\,\tq[\theta_{v}]{R}\,\vq[]{v}
    + \frac{h}{h_\mathrm{SQL}} \vq[b]{t}. \label{eq:pref}
\end{align}
Equation~\eqref{eq:pref} is equivalent to the frequency-dependent input squeezing scheme~\cite{PhysRevD.65.022002}, in which a squeezed vacuum is injected into the interferometer after pre-rotation by two filter cavities. With two available rotations, the pre-filtering scheme can be applied to detuned interferometers, as discussed in Ref.~\cite{PhysRevA.110.022601}.

\subsection{Post-Filtering}
We next consider frequency-dependent post-filtering, which is the main result of this paper. In this case, only Victor's mode is pumped, i.e., $\omega_\mr{v}=\omega_0$ and \(\vq[a,b]{t}=\{0,0\}^T\) (see Fig.~\ref{fig:Fig_modes_teleportation}b). Assuming that the operations for Alice and Bob are phase rotations,
\begin{align}
    \tq[a,b]{H} = \tq[\theta_{a,b}]{R}
    = \matr{\cos\theta_{a,b}}{-\sin\theta_{a,b}}
           {\sin\theta_{a,b}}{\cos\theta_{a,b}},
\end{align}
and that Victor's operation is the shear matrix for ponderomotive squeezing, the output becomes
\begin{align}
    \vq[]{B} \rightarrow \tq[\theta_{b}]{R}\,\tq[\theta_{a}]{R}
    \Bigl[\tq[v]{H}\,\vq[]{v} + \vq[v]{t}\,\frac{h}{h_\mathrm{SQL}}\Bigr].
\end{align}
This expression is equivalent to the post-filtering scheme described in Ref.~\cite{PhysRevD.65.022002}, which employs two filter cavities.

It should be noted that teleportation-based post-filtering does not reproduce the sensitivity curve of Ref.~\cite{PhysRevD.69.102004} because of the presence of EPR noise \(\tq{S}_{\hat{z}}\). Since this noise contributes as shot noise, its spectrum is flat and does not follow the frequency dependence of the signal, leading to degradation of the dip achieved via post-filtering. A detailed analysis of this effect is provided in a later section.

\section{Sensitivity analysis of the post-filtering approach}\label{sec.4}
In the ideal lossless case, the response of a detuned interferometer differential-mode cavity to the carrier field can be written as~\cite{PhysRevD.64.042006}
\begin{align}
    \begin{pmatrix}
        \hat{X}_1 \\
        \hat{X}_2
    \end{pmatrix}
    &= \frac{1}{\tilde{M}}
    \left(
    \begin{bmatrix}
        C_{11} & C_{12} \\
        C_{21} & C_{22}
    \end{bmatrix}
    \begin{bmatrix}
        \hat{x}_1 \\
        \hat{x}_2
    \end{bmatrix}
    + \vect{D_1}{D_2}\frac{h}{h_\mr{SQL}}
    \right),
\end{align}
where
\begin{align}
    \tilde{M} &= 1+\rho^2 e^{4i\beta}
    -2\rho e^{2i\beta}\Bigl(\cos2\phi+\tfrac{\mathcal{K}}{2}\sin 2\phi\Bigr), \\
    C_{11} &= C_{22} = (1+\rho^2)\Bigl(\cos2\phi+\tfrac{\mathcal{K}}{2}\sin 2\phi\Bigr)-2\rho\cos 2\beta, \\
    C_{12} &= -\tau^2 (\sin 2\phi + \mathcal{K} \sin^2\phi), \\
    C_{21} &= \tau^2 (\sin 2\phi - \mathcal{K} \cos^2\phi), \\
    D_1 &= -(1+\rho e^{2i\beta})\sin\phi, \\
    D_2 &= (-1+\rho e^{2i\beta})\cos \phi.
\end{align}
The parameters are
\begin{align}
    \beta &\equiv \arctan(\Omega/\gamma), \\
    \mathcal{K} &\equiv \frac{2(I_0/I_\mr{SQL})\gamma^4}{\Omega^2(\gamma^2+\Omega^2)},
\end{align}
as given in Eqs.~(2.20)–(2.24) of Ref.~\cite{PhysRevD.64.042006}.

Here, \(\gamma\) is the effective half-bandwidth, while \(\rho\) and \(\tau\) denote the amplitude reflectivity and transmissivity of the signal extraction mirror (SEM). The parameter \(\phi\) is the detuning of the signal extraction cavity (SEC), and \(I_0\) is the laser power at the beam splitter (BS). The power required to reach the standard quantum limit (SQL) at \(\Omega = \gamma\) is
\begin{align}
    I_\mr{SQL} = \frac{m L^2\gamma^4}{4\omega_0},
\end{align}
where \(m\) is the mirror mass, \(L\) is the arm cavity length, and \(\omega_0\) is the laser frequency. We do not apply the scaling law presented in Ref.~\cite{PhysRevD.67.062002}. In the limit \(I_0 \rightarrow 0\), the interferometer acts as a passive optical filter.

In the post-filtering approach, only Victor's mode is pumped, while Alice's and Bob's fields undergo phase rotations represented by the matrices \( e^{i\beta_\mr{a,b}}\tq[\theta_{a,b}]{R} \), where the overall phase factors are irrelevant to the filtering process. The interferometer parameters corresponding to each field are labeled with the subscripts \(v, a,\) and \(b\).

According to Ref.~\cite{PhysRevD.69.102004}, the input–output relation for a conventional detector is
\begin{align}
    \vect{\hat{X}_1}{\hat{X}_2}
    = \frac{1}{\tilde{M}}
    \matr{C_{11}^\xi}{C_{12}^\xi}{C_{21}^\xi}{C_{22}^\xi}
    \vect{\hat{x}_1}{\hat{x}_2}
    + \vect{D_1}{D_2}\frac{h}{h_\mr{SQL}},
\end{align}
where
\begin{align}
    \matr{C_{11}^\xi}{C_{12}^\xi}{C_{21}^\xi}{C_{22}^\xi}
    &= \matr{
        C_{11}\cos\xi + C_{12}\sin\xi
    }{
        C_{12}\cos\xi - C_{11}\sin\xi
    }{
        C_{21}\cos\xi + C_{22}\sin\xi
    }{
        C_{22}\cos\xi - C_{21}\sin\xi
    }.
\end{align}
Here \(\xi\) corresponds to \(\alpha\) in Ref.~\cite{PhysRevD.69.102004}.

When \(\hat{x}_1\) is the squeezed quadrature, i.e., \(S_{\hat{x}_1\hat{x}_1}=e^{-2r}\) and \(S_{\hat{x}_2\hat{x}_2}=e^{2r}\), the strain sensitivity (see Eq.~(64) in Ref.~\cite{PhysRevD.69.102004}) is
\begin{widetext}

\begin{align}
    S_h(\Omega) = h_\mr{SQL}^2
    \frac{
        e^{-2r}(C_{11}^\xi \cos\zeta + C_{21}^\xi \sin\zeta)^2
        + e^{2r}(C_{12}^\xi \cos\zeta + C_{22}^\xi \sin\zeta)^2
    }{|D_1\cos\zeta+D_2\sin\zeta|}. \label{eq:S_h_nofilter}
\end{align}
    
\end{widetext}
The goal of both pre-filtering and post-filtering is to eliminate the anti-squeezed component,
\begin{align}
    C_{12}^\xi \cos\zeta + C_{22}^\xi \sin\zeta = 0,
\end{align}
so that the sensitivity becomes
\begin{align}
    S_h(\Omega) = h_\mr{SQL}^2
    \frac{e^{-2r}(C_{11}^\xi \sin\zeta + C_{21}^\xi \cos\zeta)^2}
    {|D_1\cos\zeta+D_2\sin\zeta|}. \label{eq:S_h_with_filter}
\end{align}

For frequency-dependent input squeezing, \(\xi\) varies with frequency, \(\xi(\Omega)\). When detecting the amplitude quadrature \(\hat{X}_1\) (\(\zeta=0\)), \(\xi(\Omega)\) is chosen such that
\begin{align}
   C_{12}^\xi =  C_{12}\cos\xi - C_{11}\sin\xi = 0,
\end{align}
which gives
\begin{align}
    \xi_\mr{opt}(\Omega) = \arctan\frac{C_{12}}{C_{11}}.
\end{align}
For detuned interferometers, this angle is realized with two filter cavities such that
\begin{align}
    \theta_1(\Omega) + \theta_2(\Omega) = \xi_\mr{opt}(\Omega).
\end{align}

Post-filtering instead requires
\begin{align}
    C_{12}\cos\zeta + C_{22}\sin\zeta = 0,
\end{align}
when \(\xi=0\). The optimal angle is then
\begin{align}
    \zeta_\mr{opt}(\Omega) = -\arccot\frac{C_{22}}{C_{12}}, \label{eq:zeta_opt}
\end{align}
and, since \(C_{11}=C_{22}\), we obtain
\begin{align}
    \xi_\mr{opt} = \zeta_\mr{opt} + \frac{\pi}{2}.
\end{align}
Thus, the optimal filters for pre-filtering with \(\zeta=0\) can also be applied to post-filtering with \(\xi=0\), by adding an offset of \(\pi/2\).

Next we show that teleportation-based filtering achieves the same effect as the physical filter cavities. The measurement outcomes from the Bell and homodyne detection are
\begin{align}
\alpha_1 = \tfrac{1}{\sqrt{2}}\Bigl\{ & \tfrac{1}{\tilde{M}}\bigl(C_{11}\hat{v}_1 + C_{12}\hat{v}_2 + D_1\,\tfrac{h}{h_\mr{SQL}}\bigr) \notag \\
& - e^{i\beta_\mr{a}}(\cos\theta_\mr{a}\hat{a}_1 - \sin\theta_\mr{a}\hat{a}_2) \Bigr\}, \\
\alpha_2 = \tfrac{1}{\sqrt{2}}\Bigl\{ & \tfrac{1}{\tilde{M}}\bigl(C_{21}\hat{v}_1 + C_{22}\hat{v}_2 + D_2\,\tfrac{h}{h_\mr{SQL}}\bigr) \notag \\
& + e^{i\beta_\mr{a}}(\sin\theta_\mr{a}\hat{a}_1 + \cos\theta_\mr{a}\hat{a}_2) \Bigr\}, \\
    \hat{B}_2 &= e^{i\beta_\mr{b}}(\sin\theta_\mr{b}\hat{b}_1 + \cos\theta_\mr{b}\hat{b}_2).
\end{align}

By combining the data with filter gains \((g_1,g_2)\), such that \(\hat{B}_2^g=\hat{B}_2-g_1\hat{\alpha}_1-g_2\hat{\alpha}_2\), we obtain the noise spectrum (ignoring signal terms):
\begin{align}
    S_{\hat{B}_2^g\hat{B}_2^g} &= S_{\hat{B}_2\hat{B}_2}
    + |g_1|^2 S_{\hat{\alpha}_1\hat{\alpha}_1}
    + |g_2|^2 S_{\hat{\alpha}_2\hat{\alpha}_2} \notag \\
    &- g_1^*S_{\hat{B}_2\hat{\alpha}_1}
    - g_1 S_{\hat{\alpha}_1\hat{B}_2}
    - g_2^*S_{\hat{B}_2\hat{\alpha}_2}
    - g_2 S_{\hat{\alpha}_2\hat{B}_2} \notag \\
    &+ g_1 g_2^* S_{\hat{\alpha}_1\hat{\alpha}_2}
    + g_1^* g_2 S_{\hat{\alpha}_2\hat{\alpha}_1}, \label{eq.Spectrum}
\end{align}
where
\begin{align*}
    S_{\hat{B}_2\hat{B}_2} &= \cosh2r, \\
    S_{\hat{\alpha}_1\hat{\alpha}_1} &= \tfrac{1}{2}\!\left[\tfrac{e^{-2r_\mr{v}}\,C_{11}^2 + e^{2r_\mr{v}}\,C_{12}^2}{|\tilde{M}|^2} + \cosh 2r\right], \\
    S_{\hat{\alpha}_2\hat{\alpha}_2} &= \tfrac{1}{2}\!\left[\tfrac{e^{-2r_\mr{v}}\,C_{21}^2 + e^{2r_\mr{v}}\,C_{22}^2}{|\tilde{M}|^2} + \cosh 2r\right], \\
    S_{\hat{\alpha}_1\hat{\alpha}_2} &= \tfrac{1}{2}\tfrac{e^{-2r_\mr{v}}\,C_{11}C_{21} + e^{2r_\mr{v}}\,C_{12}C_{22}}{|\tilde{M}|^2}, \\
    S_{\hat{B}_2\hat{\alpha}_1} &= S^*_{\hat{\alpha}_1\hat{B}_2}
    = -\tfrac{e^{i(\beta_\mr{b}-\beta_\mr{a})}\sin(\theta_\mr{b}+\theta_\mr{a})\sinh2r}{\sqrt{2}}, \\
    S_{\hat{B}_2\hat{\alpha}_2} &= S^*_{\hat{\alpha}_2\hat{B}_2}
    = -\tfrac{e^{i(\beta_\mr{b}-\beta_\mr{a})}\cos(\theta_\mr{b}+\theta_\mr{a})\sinh2r}{\sqrt{2}}.
\end{align*}
Here we explicitly defined the squeezing level of Victor as $r_\mr{v}$ and that of the EPR entanglement as $r$.
The noise spectrum \(S_{\hat{B}_2^g\hat{B}_2^g}\) reaches its minimum when \(g_1\) and \(g_2\) are chosen as Wiener filters,
\begin{align}
g_1 &= \frac{S_{\hat{B}_2\hat{\alpha}_1} S_{\hat{\alpha}_2\hat{\alpha}_2}
       - S_{\hat{\alpha}_2\hat{\alpha}_1} S_{\hat{B}_2\hat{\alpha}_2}}
      {S_{\hat{\alpha}_1\hat{\alpha}_1} S_{\hat{\alpha}_2\hat{\alpha}_2}
       - \left|S_{\hat{\alpha}_1\hat{\alpha}_2}\right|^2}, \notag\\
g_2 &= \frac{S_{\hat{B}_2\hat{\alpha}_2} S_{\hat{\alpha}_1\hat{\alpha}_1}
       - S_{\hat{\alpha}_1\hat{\alpha}_2} S_{\hat{B}_2\hat{\alpha}_1}}
      {S_{\hat{\alpha}_1\hat{\alpha}_1} S_{\hat{\alpha}_2\hat{\alpha}_2}
       - \left|S_{\hat{\alpha}_1\hat{\alpha}_2}\right|^2}.
\end{align}
Here we used the relation \(\theta_b+\theta_a=\zeta\). The gravitational-wave signal is transferred through filtering with transfer function
\begin{align}
    \tau_h = \frac{g_1 D_1 + g_2 D_2}{\sqrt{2}\tilde{M}}.
\end{align}
The final strain sensitivity is
\begin{align}
     S_h = h_\mr{SQL}^2\frac{S_{\hat{B}_2^g\hat{B}_2^g}}{|\tau_h|^2}. \label{eq:SNR}
\end{align}

The noise spectrum is given by:
\newcommand{\DOne}{C_{11}^2 + C_{21}^2} 
\newcommand{\T}{\left( C_{11} + C_{21}\tan\zeta \right)^2}
\begin{align}
S_{\hat{B}_2^g\hat{B}_2^g} = \frac{N}{D}, \label{eq:Strain} 
\end{align}
with
\begin{widetext}

\begin{align}
N &= \DOne\,\cosh^2 2r  
  - \sinh^2 2r  \bigl( C_{11}^2\sin^2\zeta - C_{11}C_{21}\sin2\zeta + C_{21}^2\cos^2\zeta \bigr) \notag\\
  &\quad + e^{2r_v}\Bigl[ e^{2r_v}\, \sec^2\zeta\, C_{22}^2 
      + \cosh2r \bigl( 1 + C_{22}^2\,\T \bigr) \Bigr], \\
D &= \cosh2r\, \DOne 
    + e^{4r_v}\,\cosh2r\, \sec^2\zeta\, C_{22}^2 \notag\\
  &\quad + e^{2r_v}\bigl( \cosh^2 2r + C_{22}^2\,\T \bigr). \label{eq:D}
\end{align}
    
\end{widetext}
In the limit \(r\rightarrow\infty\), the filters and strain sensitivity reduce to
\begin{align}
    g_1 &= -\sqrt{2}\,e^{i(\beta_\mr{b}-\beta_\mr{a})}\cos \zeta, \\
    g_2 &= -\sqrt{2}\,e^{i(\beta_\mr{b}-\beta_\mr{a})}\sin \zeta, \\
    S_h^{(0)} &= \tfrac{h_\mr{SQL}^2}{2}\,
    \frac{e^{-2r_\mr{v}}(C_{11}\sin\zeta+C_{21}\cos\zeta)^2}{|D_1 \sin\zeta+D_2\cos\zeta|^2}. \label{eq:S_h_teleported}
\end{align}
Equation~\eqref{eq:S_h_teleported} is equivalent to Eq.~\eqref{eq:S_h_with_filter}.

Our main interest lies in the signal-to-noise ratio of Eq.~\eqref{eq:SNR}, rather than the spectrum of Eq.~\eqref{eq.Spectrum}. Equation~\eqref{eq:SNR} is not directly differentiable with respect to \(g_1\) and \(g_2\). To find its minimum, one must expand all complex functions into real and imaginary parts and compute the derivative of \(S_h\). This procedure is mathematically rigorous but considerably complicates the analysis.

In the limit \(r\rightarrow \infty\), Eq.~\eqref{eq:Strain} converges to Eq.~(68) of Ref.~\cite{PhysRevD.69.102004}. We therefore regard the filters \(g_1\) and \(g_2\) as suboptimal filters.

\section{Sensitivity in the Einstein Telescope}\label{sec.5}
In this section, we apply the post-filtering scheme to the low-frequency detector within the Einstein Telescope xylophone configuration (ETLF)~\cite{Hild_2011}. The ETLF is designed with a detuned SEC to enhance sensitivity at 8\,Hz through the optical spring effect. 

For detuned interferometers, at least two filter cavities are required to compensate for the frequency-dependent squeezing angle rotation caused by the optical spring and optical resonances. Since the target sensitivity of the ETLF is extremely high below about 20\,Hz, the planned length of each filter cavity is several kilometers. In the Einstein Telescope, the construction cost of such long filter cavities is considerable, both due to their length and the underground site of the detectors. 

The parameters of the ETLF are summarized in Table~\ref{Tab1}. In the following, we analyze the sensitivity using the post-filtering approach under both lossless (Sec.~\ref{s.sec.lossless}) and lossy (Sec.~\ref{s.sec.lossy}) conditions, and compare it with the pre-filtering approach.

\begin{table}[h]
\centering
\begin{tabular}{ p{1cm}||p{4.cm}|p{2cm}}
 \hline
 \multicolumn{3}{c}{Parameter List} \\
 \hline
    $\lambda$ & Carrier wavelength & 1550 nm \\ \hline
  $T_\mathrm{arm}$ & ITM power transmittance & 7000 ppm \\ \hline
  $T_\mathrm{SEM}$ & SEM power transmittance & 20 \% \\ \hline
  $m$ & Mirror mass & 211 kg \\ \hline
  $I_0$ & Power at BS & 63 W  \\ \hline
  $\phi$ & Detuning of the SEC & 0.75 rad \\ \hline
  $L_\mathrm{arm}$ & Arm length & 10 km \footnotemark[1]   \\ \hline
  $L_\mathrm{SEC}$ & SEC length & 100 m \footnotemark[1]  \\ \hline
\end{tabular}
\footnotetext[1]{For the QT scheme, $L_\mathrm{arm}=6451612903\lambda$ and $L_\mathrm{SEC}=64516129\lambda$, where $\lambda$ is the wavelength of the main laser.}
\caption{\label{Tab1} Parameters for ETLF~\cite{Hild_2011}}
\end{table}

\begin{table*}[h]
\caption{Parameters for squeezing in ETLF}\label{tab1}%
\begin{tabular}{p{5.5cm}|p{4.cm}<{\centering}} \hline
    Arm round trip loss & 45 ppm \\ \hline
    SEC loss & 1000 ppm \\ \hline
    Injection loss & 4 \% \\ \hline
    Readout loss & 3 \% \\ \hline
    Effective FC loss & $\sim 6$ ppm/km \\ \hline
    Squeezer noise RMS\footnotemark[1] & 10 mrad \\ \hline
    Local oscillator RMS & 10 mrad \\ \hline
    SEC length RMS & 1 pm \\ \hline
    Bandwidth/detuning of the first FC & 4.27/19.54 Hz \\ \hline
    Bandwidth/detuning of the second FC & 1.64/-7.62 Hz \\ \hline
    Detuning of Alice $\omega_\mr{a}-\omega_\mr{0}$ & $768\ \mathrm{kHz} + 11\ \mathrm{FSR}_\mathrm{SEC}$ \\ \hline
    Detuning of Victor or Bob $\omega_{\mr{v,b}}-\omega_\mr{0}$  & $655\ \mathrm{kHz} +1059\ \mathrm{FSR}_\mathrm{SEC}$ \\ \hline
    Squeezing level & \multicolumn{1}{c}{15 dB} \\ \hline
\end{tabular}
\footnotetext[1]{Root mean square}
\end{table*}

\subsection{Lossless case} \label{s.sec.lossless}
Before calculating the strain sensitivity, we decompose the noise contributions from Victor and the EPR pair, which provides a useful interpretation of the scheme. The strain sensitivity can be written as the sum of these contributions,
\begin{align}
    S_h \sim S_h^{(0)} + S_h^{\mr{EPR}},
\end{align}
where
\begin{align}
    S_h^{\mr{EPR}} = e^{-2r}\left.\frac{\mr{d}S_h}{\mr{d}e^{-2r}}\right|_{r\rightarrow\infty}.
\end{align}
Here, \( S_h^{\mr{EPR}} \) corresponds to the first term in Eq.~\eqref{eq:B}. 

\begin{figure}
    \centering
    \includegraphics[width=1\linewidth]{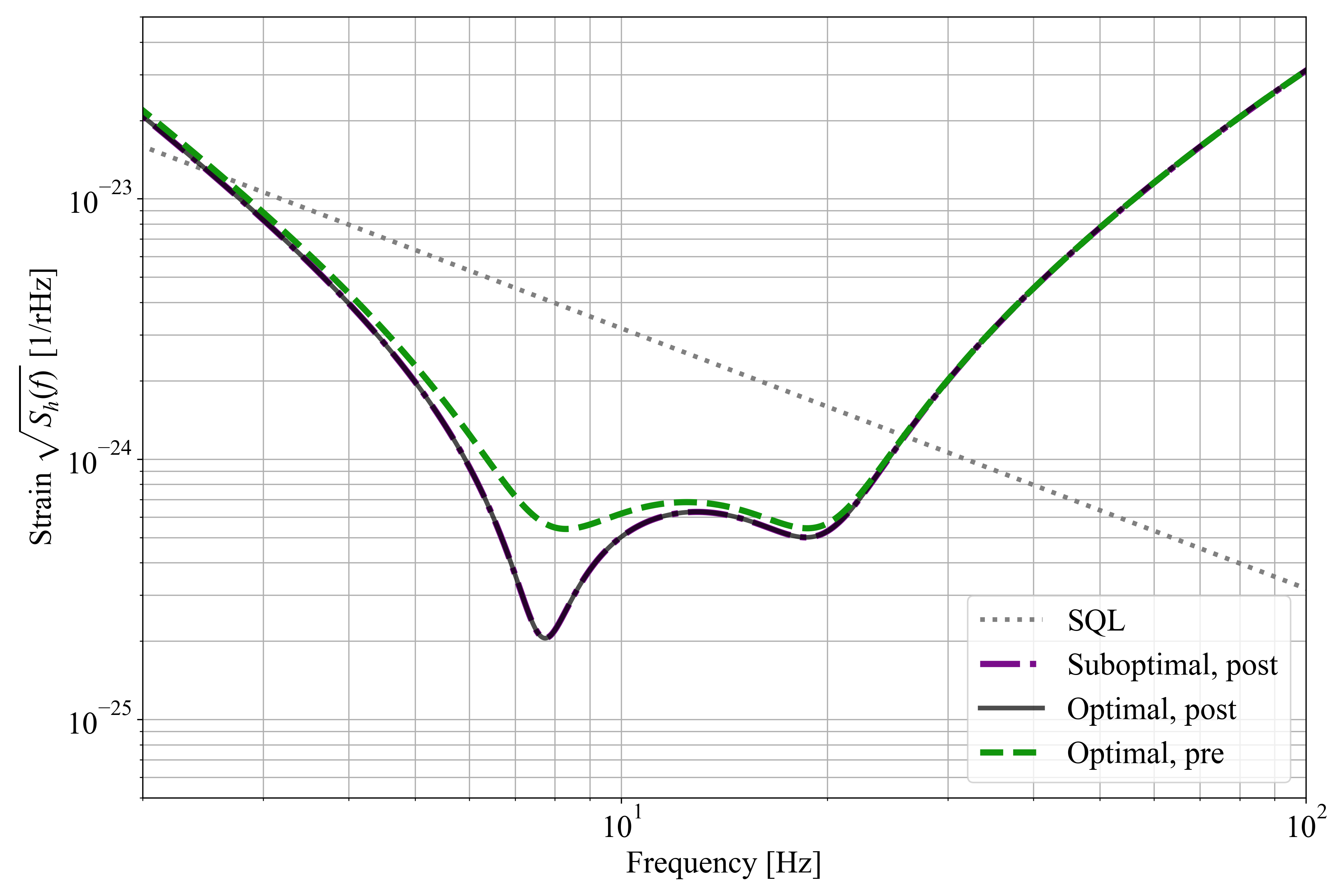}
    \caption{Noise contribution in the optimal (solid black) and suboptimal (dash-dotted purple) readout schemes in the limit of $F\rightarrow1$. The green dashed curve is the optimal pre-filtering result. The dotted line indicates the SQL.}
    \label{fig:fig_sens_Suboptimal_optimal_victor}
\end{figure}

Figure~\ref{fig:fig_sens_Suboptimal_optimal_victor} shows \( S_h^{(0)} \) for pre-filtering and post-filtering in the limit of perfect teleportation fidelity ($F\rightarrow1$). For post-filtering, both suboptimal (solid black) and optimal (dash-dotted purple) cases are shown. Under the present parameter settings, the suboptimal case is close to the optimal case (see Appendix~\ref{sec.a.1}). We therefore focus on the suboptimal case, where the optimal phase rotation can be realized with passive cavities. The optimal pre-filtering result is also shown as the dashed green curve. These results indicate that, in the lossless and perfect-fidelity limit, post-filtering outperforms pre-filtering near the optical spring frequency.

\begin{figure}
    \centering
    \includegraphics[width=1\linewidth]{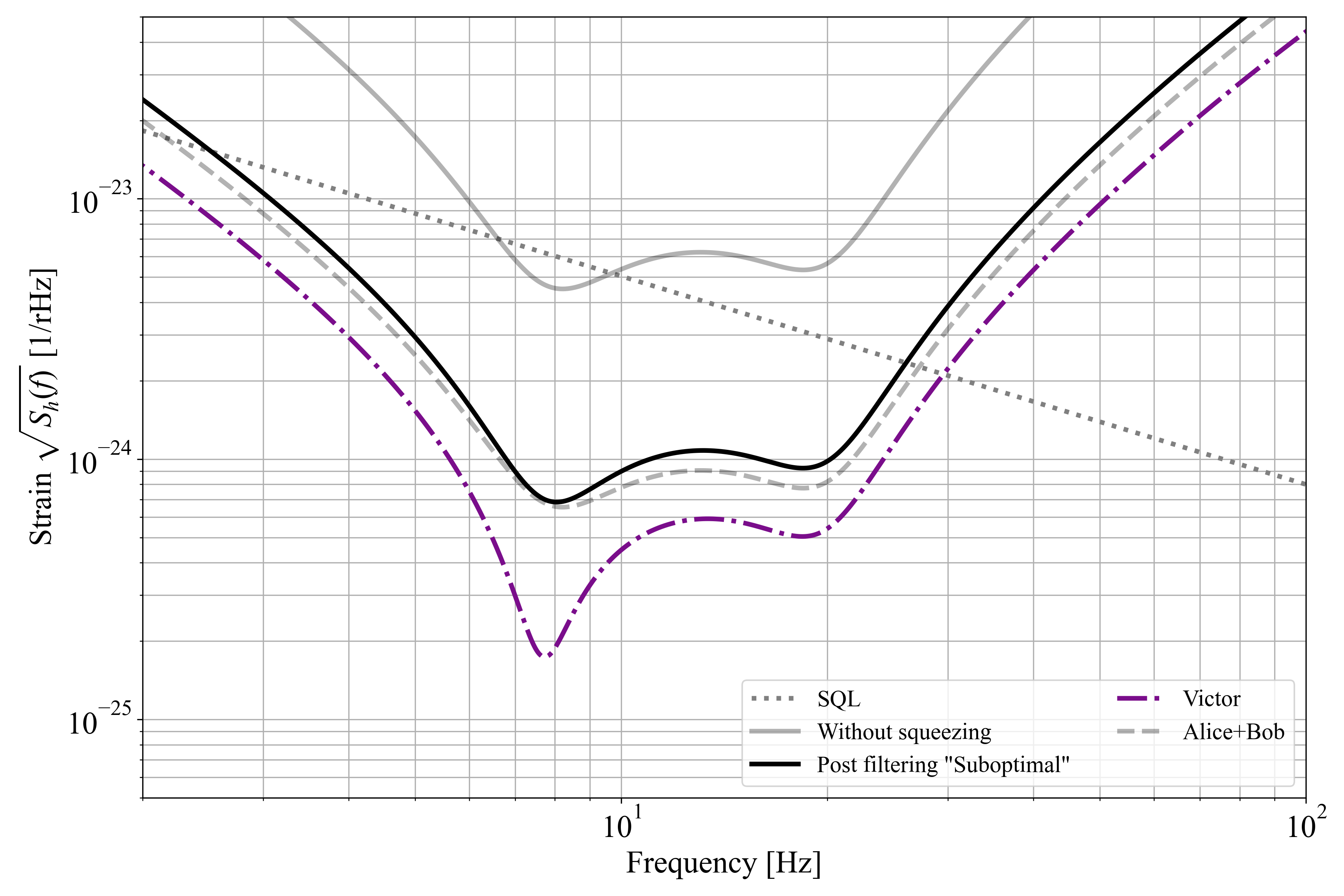}
    \caption{Noise budget of the post-filtering scheme in the lossless limit. The purple dash-dotted curve represents Victor, the gray dashed curve corresponds to the EPR pair, the solid black curve shows the total noise, and the solid gray curve shows the case without squeezing.}
    \label{fig:fig_sens_Suboptimal}
\end{figure}

Figure~\ref{fig:fig_sens_Suboptimal} shows the noise budget in the lossless but non-perfect fidelity case, calculated using Eqs.~\eqref{eq:Strain}–\eqref{eq:D}. The squeezing parameters are listed in Table~\ref{tab1}. The purple dash-dotted and gray dashed curves represent the contributions from Victor (\(\sim S_h^{(0)}\)) and the EPR pair (\(\sim S_h^{\mr{EPR}}\)), respectively, while the solid black curve shows their sum. Both Victor and the EPR pair are assumed to have a squeezing level of 15\,dB. 

The results show that the sensitivity enhancement achieved through filtering Victor is degraded by the EPR noise. This occurs because the EPR noise behaves as white noise and does not exhibit the dip structure associated with the signal response. In the limit \(r\rightarrow\infty\), the EPR noise scales as \(e^{-2r}\), leaving only the contribution from Victor. In principle, this suggests that stronger squeezing could further improve sensitivity. In practice, however, imperfections limit the achievable improvement, as discussed in the next section.

\begin{figure*}[htbp]
    \centering
    \begin{minipage}{0.48\linewidth}
        \centering
        \includegraphics[width=\linewidth]{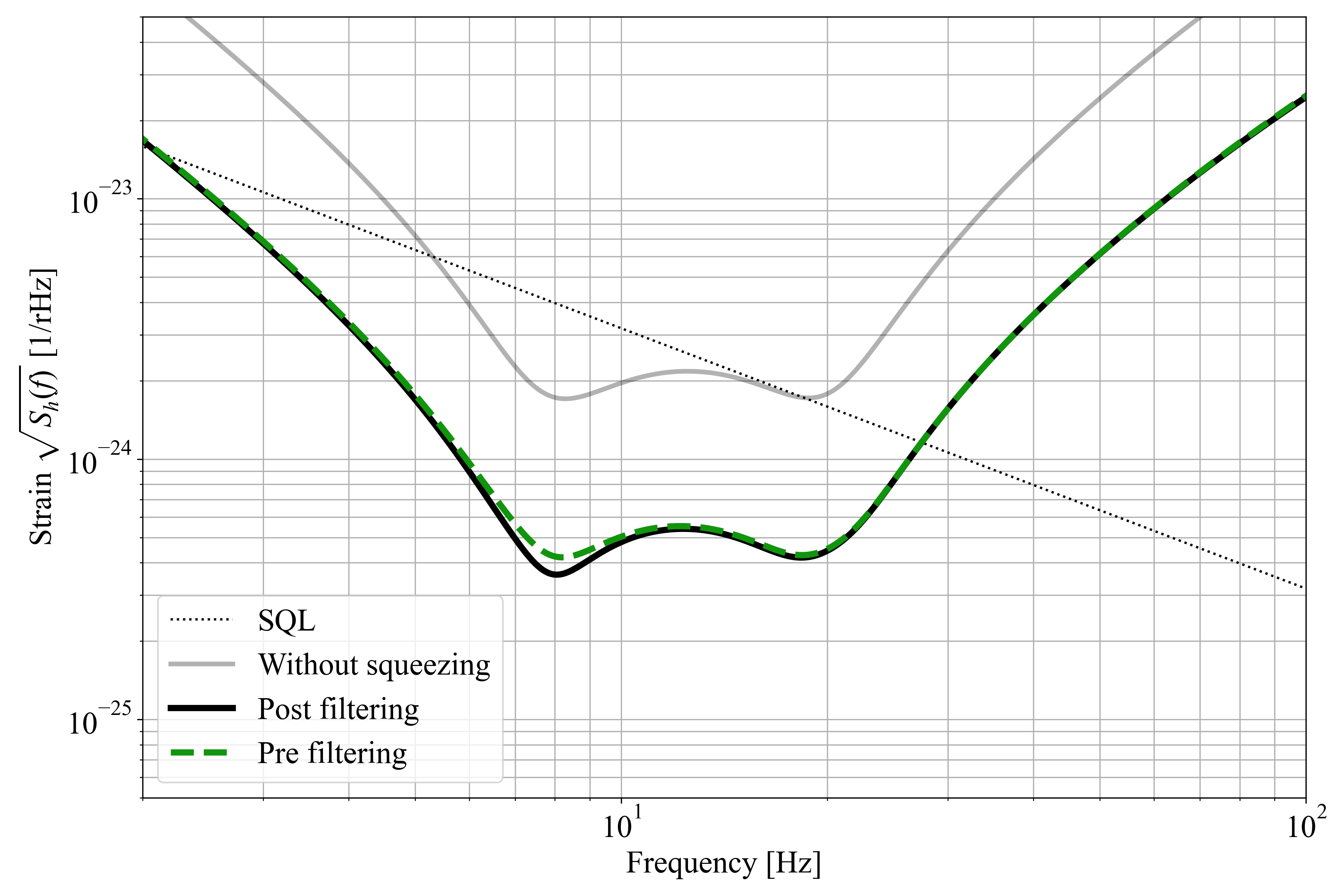}
    \end{minipage}
    \hfill
    \begin{minipage}{0.48\linewidth}
        \centering
        \includegraphics[width=\linewidth]{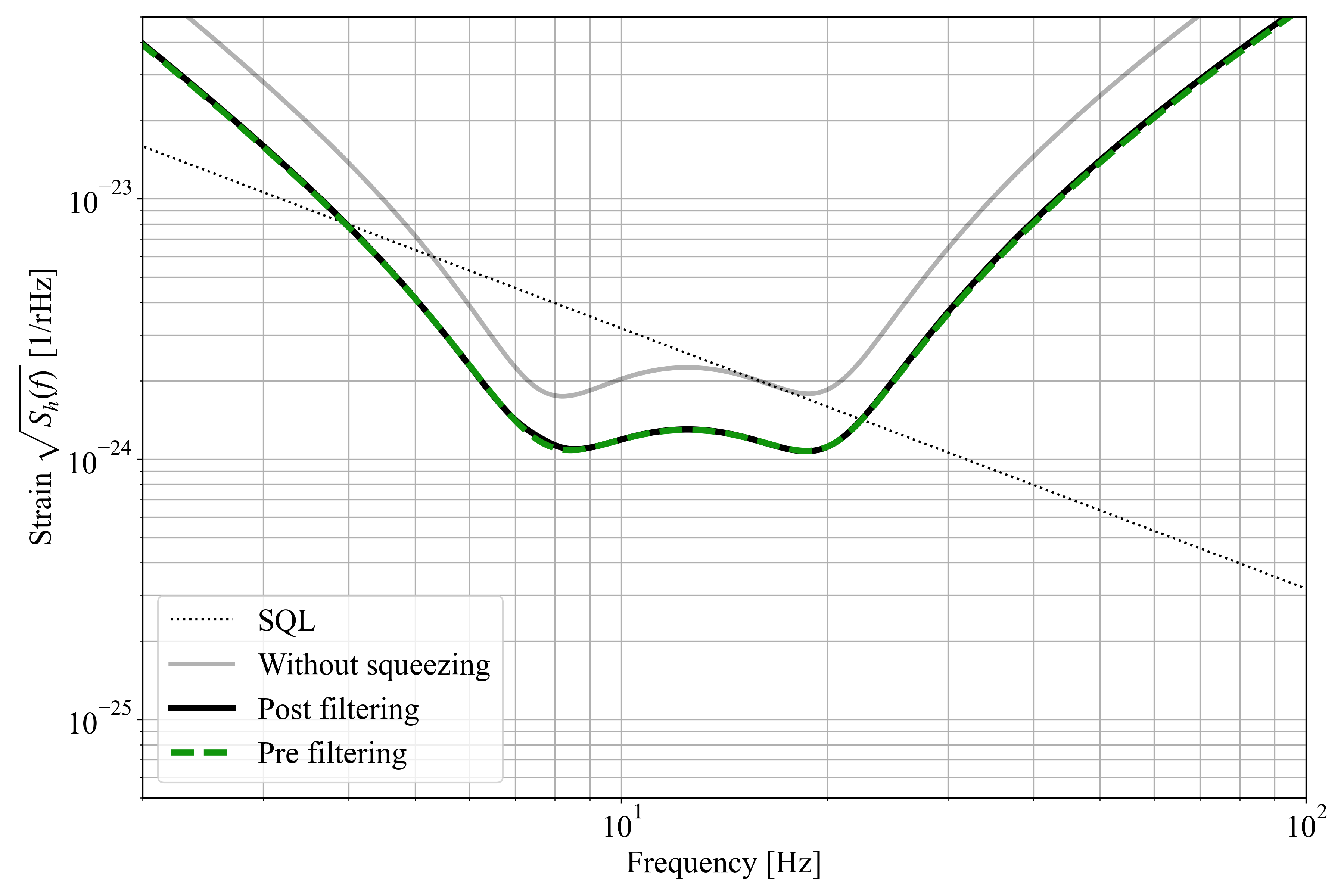}
    \end{minipage}
    \caption{Comparison of pre-filtering and post-filtering. Left: lossless case. Right: lossy detectors.}
    \label{fig:combined}
\end{figure*}

A comparison between post-filtering and pre-filtering is shown in the left panel of Fig.~\ref{fig:combined}. The post-filtering scheme exhibits a small sensitivity enhancement near the optical spring frequency (\(\sim 8\) Hz). This enhancement arises from frequency-dependent post-filtering of Victor, operating through the same mechanism as described in Ref.~\cite{PhysRevD.69.102004}.

\subsection{Lossy case}\label{s.sec.lossy}
We now include losses and imperfections in the analysis. The loss parameters, summarized in Table~\ref{tab1}, are the same as those in Ref.~\cite{PhysRevA.110.022601}. 

\begin{figure*}[htbp]
    \centering
    \begin{minipage}{0.48\linewidth}
        \centering
        \includegraphics[width=\linewidth]{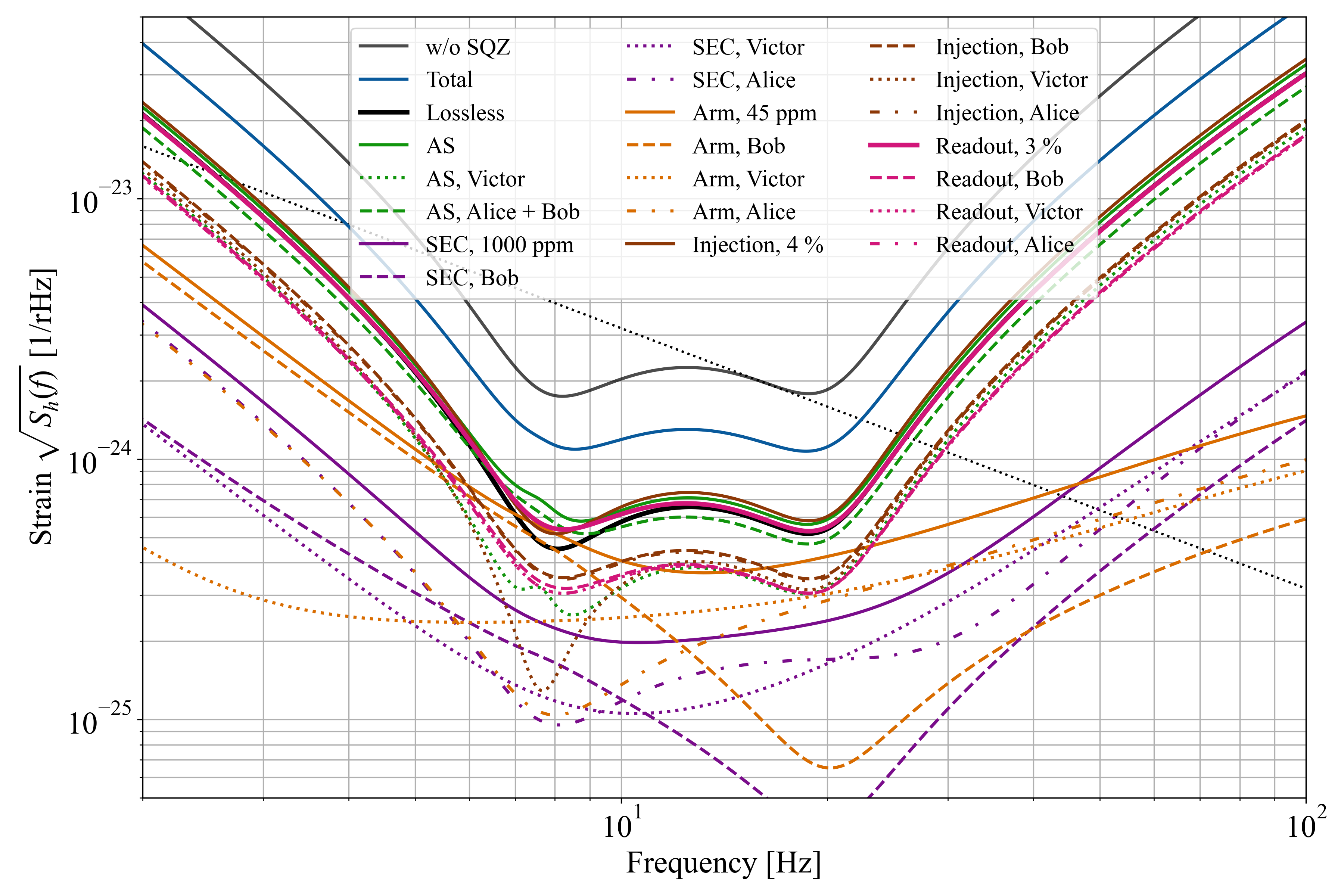}
    \end{minipage}
    \hfill
    \begin{minipage}{0.48\linewidth}
        \centering
        \includegraphics[width=\linewidth]{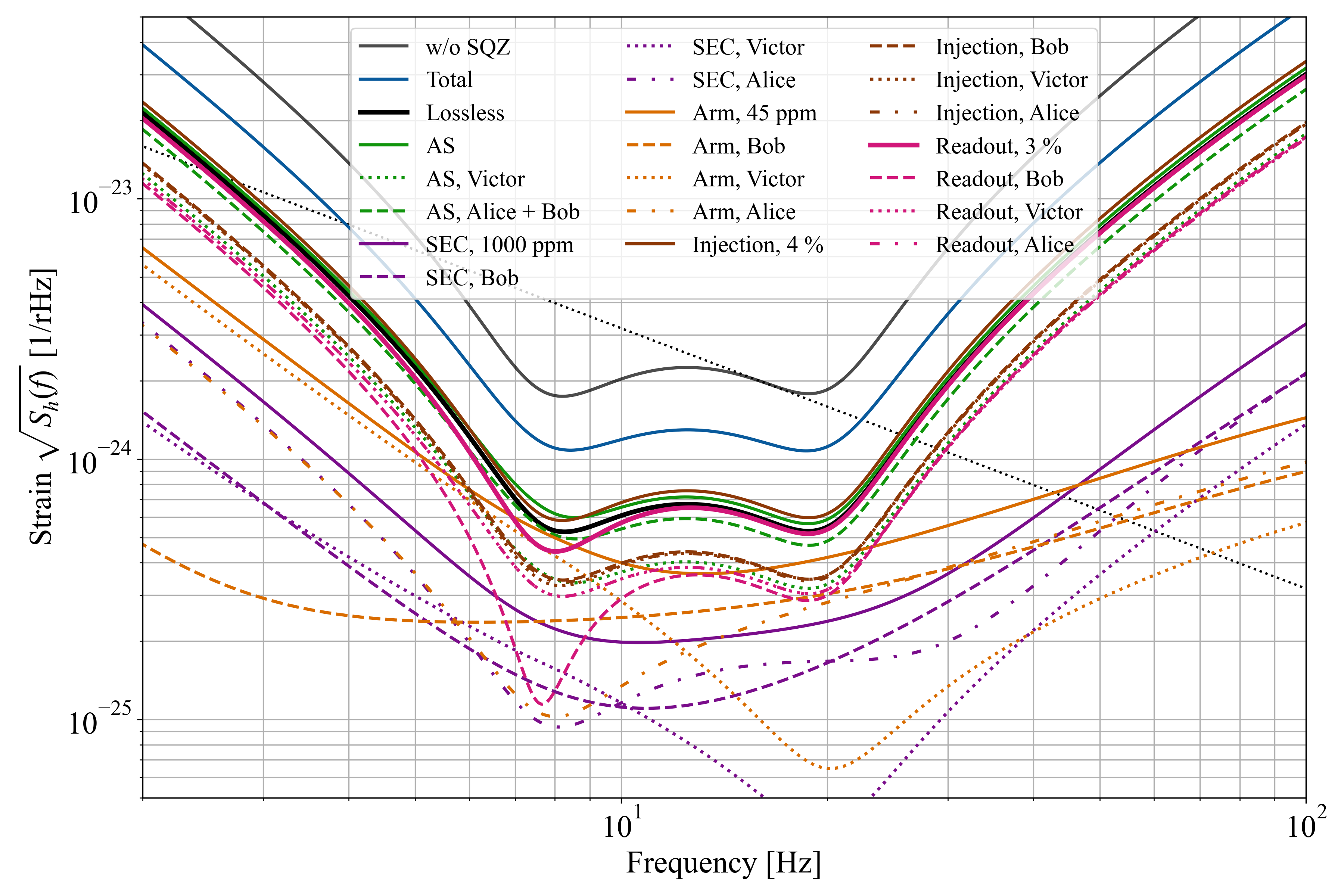}
    \end{minipage}
    \caption{Noise budget for lossy post-filtered (left) and pre-filtered (right) detectors.}
    \label{fig:combined_budget}
\end{figure*}

We assume the same squeezing level for both schemes: 15\,dB applied to Victor and the EPR pair. The full noise budgets for the post-filtering and pre-filtering schemes are shown in Fig.~\ref{fig:combined_budget}. The contributions from loss-induced vacuum noise are further decomposed into frequency channels corresponding to Victor, Alice, and Bob. The solid curve shows the total contribution for each channel, while the dashed, dotted, and dot-dashed curves show individual components. 

For both schemes, injection and readout losses dominate the noise budget. In the post-filtering case, the readout loss contribution from Bob is larger near 8\,Hz than in the pre-filtering case. Because Bob performs phase rotation in post-filtering, the sensitivity is more affected by readout loss. The post-filtering scheme is also more sensitive to phase misalignment, which leads to leakage of the anti-squeezed quadrature of the input vacuum (labeled “AS”) into the detected quadrature. This effect appears in the left panel of Fig.~\ref{fig:combined_budget} as a bump in the green curves, but is not clearly visible in the right panel.

Finally, we overlay the total sensitivity curves for both schemes in the right panel of Fig.~\ref{fig:combined}, using the same color scheme as in the left panel. The two curves are nearly identical, indicating that teleportation-based post-filtering achieves almost the same sensitivity as pre-filtering. A more detailed quantitative comparison is presented in the next section.

\section{Signal-to-noise ratio for compact binary coalescence}\label{sec.6}
We numerically evaluate the detector performance in terms of the signal-to-noise ratio (SNR) for inspiraling compact binary systems. The matched-filter SNR is defined as
\begin{align}
    \rho^2 = \int_0^{\Omega_\mr{max}}\frac{\Omega^{-7/3}}{S_h(\Omega)} \frac{\mr{d}\Omega}{2\pi},
\end{align}
where \( \Omega_\mr{max} \) is the cutoff frequency of the event, which depends on the binary masses. Our primary interest is the ratio of the SNRs for the post-filtering and pre-filtering schemes,
\begin{align}
    \bar{\rho}^2 \equiv \frac{\rho^2_\mr{post}}{\rho^2_\mr{pre}}.
\end{align}
Throughout this analysis we set \( \Omega_\mr{max}=2\pi\times 1 \,\text{kHz} \).

\begin{figure*}[htbp]
    \centering
    \begin{minipage}{0.48\linewidth}
        \centering
        \includegraphics[width=\linewidth]{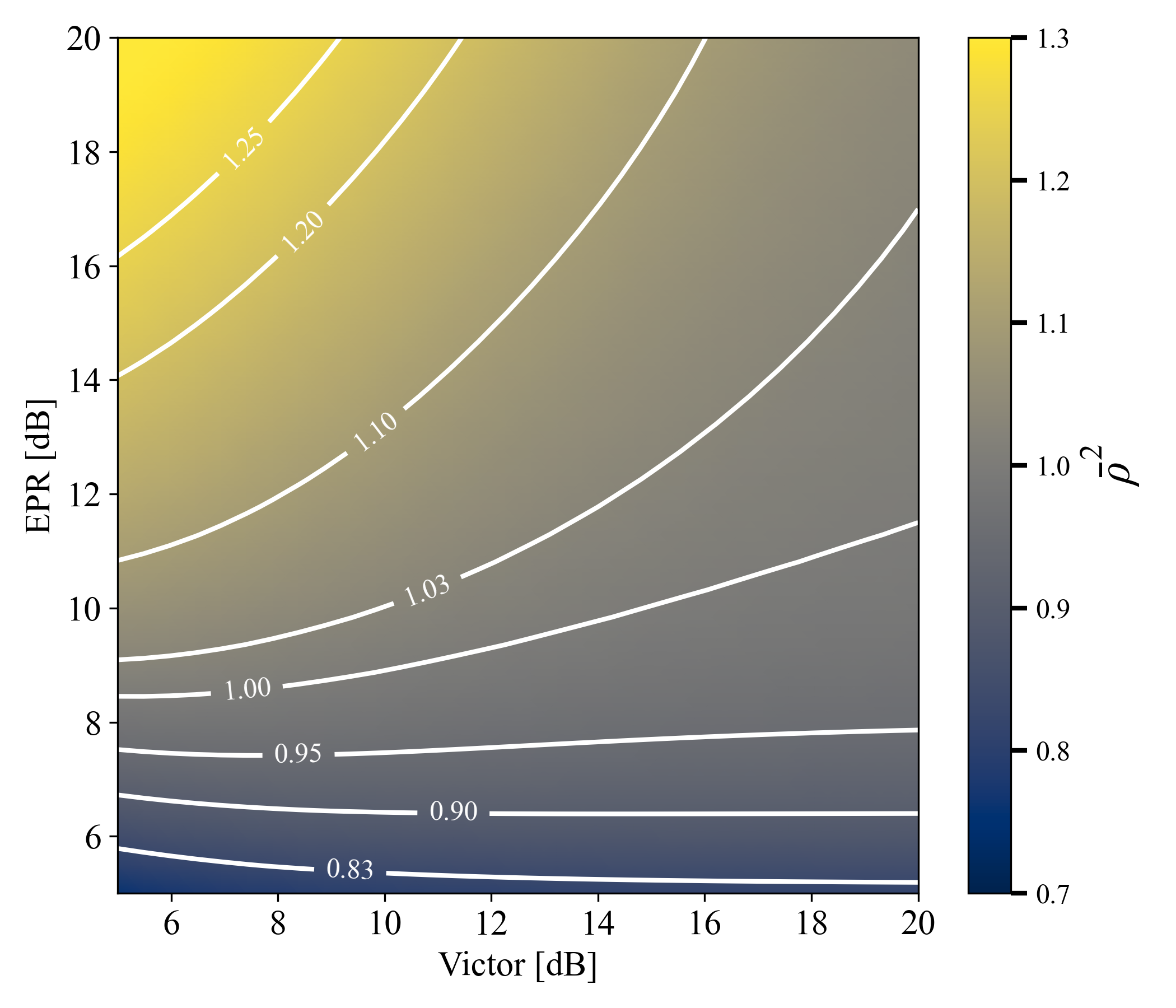}
    \end{minipage}
    \hfill
    \begin{minipage}{0.48\linewidth}
        \centering
        \includegraphics[width=\linewidth]{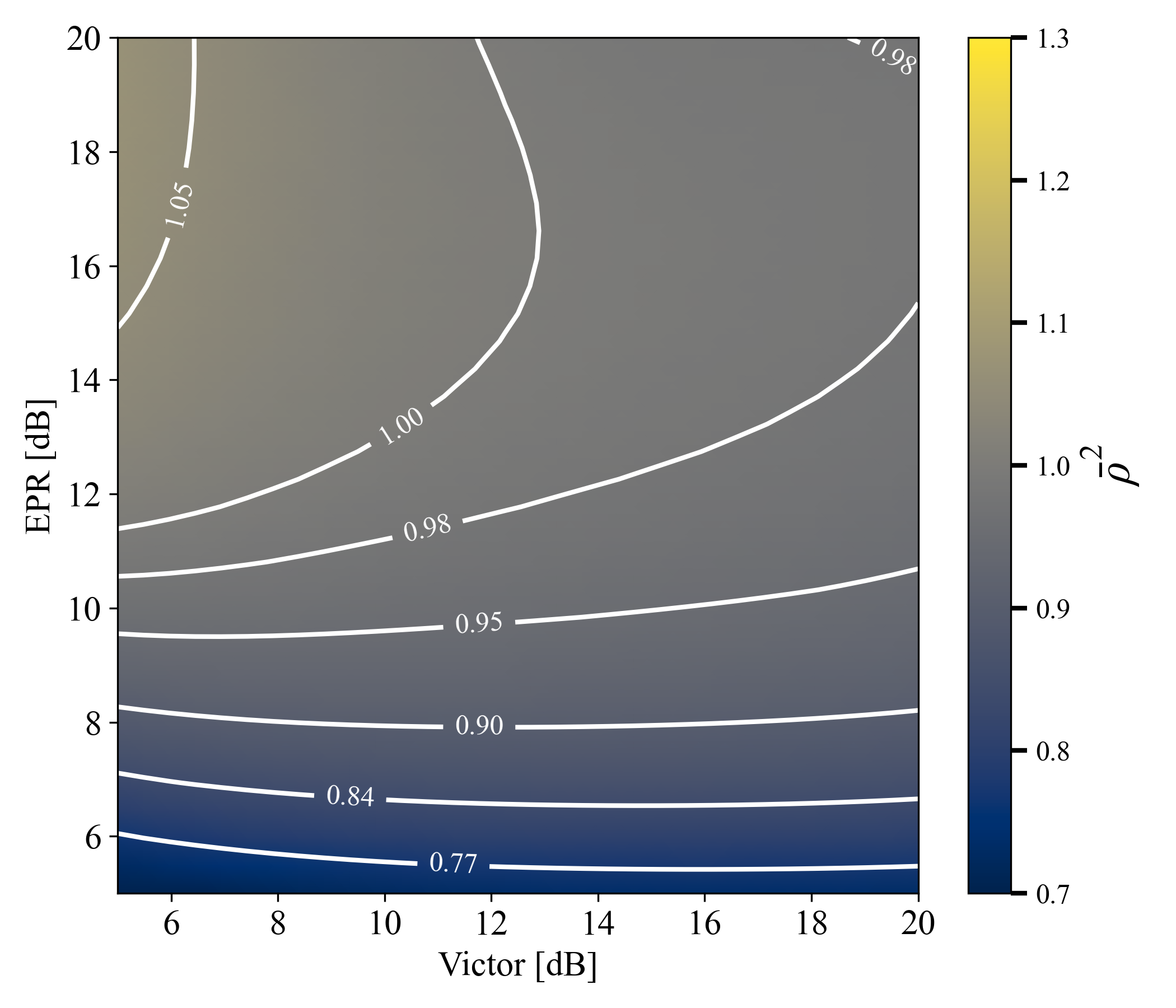}
    \end{minipage}
    \caption{Ratio $\bar{\rho}^2$ for different squeezing levels. Left: lossless case. Right: lossy case.}
    \label{fig:combined_SNR}
\end{figure*}

We evaluate this ratio for both the lossless and lossy cases. The results are shown in Fig.~\ref{fig:combined_SNR}. Different squeezing levels were used for Victor and the EPR pair ($r$ and $r_\mathrm{v}$ in Sec.~\ref{sec.3}). 
In the lossless case, post-filtering outperforms pre-filtering when the EPR pair squeezing exceeds about 8\,dB. As the squeezing level increases, the fidelity of both the squeezing and teleportation processes approaches unity, and post-filtering becomes superior to pre-filtering due to enhanced noise suppression near the optical spring frequency (see Fig.~\ref{fig:fig_sens_Suboptimal}). At lower EPR squeezing, however, the imperfection of the entanglement contributes as shot noise. Since post-filtering is more susceptible to shot noise than pre-filtering, its sensitivity degrades in this regime (see also Sec.~6.1.2 of Ref.~\cite{Danilishin_2012}).

In the lossy case, no clear advantage of post-filtering over pre-filtering is observed. Because post-filtering is more sensitive to losses and imperfections, its performance converges to that of pre-filtering, similar to conventional filter-cavity schemes. At practical squeezing levels of about 15\,dB, the ratio \(\bar{\rho}^2\) is close to unity, indicating that post-filtering can be considered as an option for future detuned detectors alongside pre-filtering.

\section{Conclusion}\label{sec.7}
In this paper we applied the principle of quantum teleportation to the frequency-dependent post-filtering scheme discussed in Ref.~\cite{PhysRevD.69.102004}. Post-filtering provides its strongest effect near the dip at the optical spring frequency, which is deeper than that obtained with pre-filtering. We showed that by changing the order of operations in the teleportation protocol, the same post-filtering as in Ref.~\cite{PhysRevD.69.102004} can be realized in the limit of perfect teleportation fidelity. In practice, however, losses and imperfections spoil the dip, making the sensitivity nearly equivalent to that of pre-filtering. A quantitative analysis based on the signal-to-noise ratio for chirp signals from compact binary coalescence shows that, in the presence of losses and imperfections, teleportation-based post-filtering is comparable to pre-filtering with realiztic squeezing levels.

\begin{acknowledgments} 
The author is grateful to Stefan Hild, Stefan Danilishin, and Teng Zhang in the ET collaboration for fruitful discussions about the simulation for the Einstein Telescope, Daniel Gould and Jonas Junker in the ANU CGA squeezer group for fruitful discussions about experimental realization and Yutaro Enomoto, Xinyao Guo, James Gardner and Yanbei Chen for discussions about theoretical calculation. Research by Y. N. is supported by JSPS Grant-in-Aid for JSPS Fellows Grant Number 23KJ0787 and JST ASPIRE (JPMJAP2320). 
\end{acknowledgments}

\appendix
\section{Optimality of filtering}\label{sec.a.1}
We show that in the high-squeezing limit the suboptimal post-filtering coincides with the optimal post-filtering. According to Ref.~\cite{PhysRevD.69.102004}, the optimal homodyne angle is obtained from the roots of the kernel function
\begin{align}
    \mathcal{F}_2 \cot^2\zeta + \mathcal{F}_1 \cot\zeta + \mathcal{F}_0 = 0, \label{eq:appendix}
\end{align}
where
\begin{widetext}
\begin{align}
    \mathcal{F}_2 &= \bigl[|M|^2 e^{-2r}+ ( C^{-r}_{11} )^2 + ( C^{r}_{12} )^2\bigr] \, \operatorname{Re}(D_1^* D_2) 
    - \bigl( C^{-r}_{11} C^{-r}_{21} + C^{r}_{12} C^{r}_{22} \bigr) |D_1|^2, \label{eq:F2} \\
    \mathcal{F}_1 &= \bigl[|M|^2 e^{-2r}+ ( C^{-r}_{11} )^2 + ( C^{r}_{12} )^2\bigr] \, |D_2|^2 
    - \bigl[ ( C^{-r}_{21} )^2 + ( C^{r}_{22} )^2 \bigr] |D_1|^2, \label{eq:F1} \\
    \mathcal{F}_0 &= \bigl( C^{-r}_{11} C^{-r}_{21} + C^{r}_{12} C^{r}_{22} \bigr) |D_2|^2 
    - \bigl[ ( C^{-r}_{21} )^2 + ( C^{r}_{22} )^2 \bigr] \operatorname{Re}(D_1^* D_2). \label{eq:F0}
\end{align}
\end{widetext}
The term \( |M|^2 e^{-2r} \) arises from the EPR noise in Eq.~\eqref{eq:B}. In the limit \(r\rightarrow\infty\), the terms with $|M|^2 e^{-2r}$, \(C_{11}^{-r}\) and \(C_{12}^{-r}\) vanish, and one solution of Eq.~\eqref{eq:appendix} is
\begin{align}
    \zeta = - \arccot \frac{C_{22}}{C_{12}},
\end{align}
which agrees with Eq.~\eqref{eq:zeta_opt}.